\documentclass[letterpaper]{article} 
\usepackage{aaai25}  
\usepackage{times}  
\usepackage{helvet}  
\usepackage{courier}  
\usepackage[hyphens]{url}  
\usepackage{graphicx} 
\usepackage{natbib}  
\usepackage{caption} 
\usepackage{algorithm}
\usepackage{algorithmic}

\usepackage{newfloat}
\usepackage{listings}
\DeclareCaptionStyle{ruled}{labelfont=normalfont,labelsep=colon,strut=off} 
\floatstyle{ruled}
\newfloat{listing}{tb}{lst}{}
\floatname{listing}{Listing}
\nocopyright

\usepackage{amsmath}
\usepackage{amssymb}
\usepackage{bm}
\usepackage{subfigure}
\usepackage{booktabs}
\usepackage{multirow}
\usepackage{lipsum} 
\usepackage[utf8]{inputenc}
\usepackage{siunitx}
\usepackage{rotating}
\usepackage{pifont}
\usepackage{amsthm}
\theoremstyle{plain}
\newtheorem{theorem}{Theorem}

\theoremstyle{definition}

\theoremstyle{remark}

\newcommand{\BEST}[1]{\textcolor{red}{#1}}
\newcommand{\SECOND}[1]{\textcolor{blue}{#1}}
\newcommand{\THIRD}[1]{\textcolor{violet}{#1}}

\newcommand{\cmark}{\ding{51}}%
\newcommand{\xmark}{\ding{55}}%
\usepackage[textsize=small]{todonotes}
\usepackage{marginnote}
 
\title{RDGCL: Reaction-Diffusion Graph Contrastive Learning for Recommendation}
\author{
    Jeongwhan Choi\equalcontrib\textsuperscript{\rm 1},
    Hyowon Wi\equalcontrib\textsuperscript{\rm 2},
    Chaejeong Lee\textsuperscript{\rm 1},
    Sung-Bae Cho\textsuperscript{\rm 1},
    Dongha Lee\textsuperscript{\rm 1},
    Noseong Park\textsuperscript{\rm 2}\thanks{Noseong Park is the corresponding author.}
}
\affiliations{
    \textsuperscript{\rm 1}Yonsei University, Seoul, South Korea\\
    \textsuperscript{\rm 2}KAIST, Daejeon, South Korea\\
    \{jeongwhan.choi, chaejeong\_lee, sbcho, donalee\}@yonsei.ac.kr,
    \{hyowon.wi, noseong\}@kaist.ac.kr
}

\begin{document}

\maketitle

\begin{abstract}
Contrastive learning (CL) has emerged as a promising technique for improving recommender systems, addressing the challenge of data sparsity by using self-supervised signals from raw data. Integration of CL with graph convolutional network (GCN)-based collaborative filterings (CFs) has been explored in recommender systems. However, current CL-based recommendation models heavily rely on low-pass filters and graph augmentations. In this paper, inspired by the reaction-diffusion equation, we propose a novel CL method for recommender systems called the reaction-diffusion graph contrastive learning model (RDGCL). We design our own GCN for CF based on the equations of diffusion, i.e., low-pass filter, and reaction, i.e., high-pass filter. Our proposed CL-based training occurs between reaction and diffusion-based embeddings, so there is no need for graph augmentations. Experimental evaluation on 5 benchmark datasets demonstrates that our proposed method outperforms state-of-the-art CL-based recommendation models. By enhancing recommendation accuracy and diversity, our method brings an advancement in CL for recommender systems.
\end{abstract}

\section{Introduction}

CL is attracting much attention and is being actively researched in the field of machine learning~\cite{jaiswal2020survey,khosla2020supervised,liu2021self,chen2021intriguing}. CL enhances the user/item embedding process with the learning representation principle, which increases the similarity between positive pairs and maximizes the dissimilarity between negative pairs. CL has achieved many successes in a variety of domains, including computer vision~\cite{chen2020simple,chen2020big,he2020momentum}, natural language processing~\cite{radford2021learning,chuang2022diffcse,bayer2022survey} and graph data~\cite{qiu2020gcc,hassani2020contrastive,zhu2021GCA,zhu2020deep}. In the field of recommender systems, recent CF methods are mostly based on it~\cite{Wu2021SGL,yu2022SimGCL,yu2022xsimgcl,cai2023lightgcl,xu2023simdcl,li2023sgccl,jing2023survey}.

The integration of CL with GCNs has great potential in tackling the data sparsity problem in recommender systems~\cite{Wu2021SGL,jing2023survey}. GCN-based CF methods excel at capturing complex dependencies and interactions among entities in graph-structured data, making them suitable for modeling user-item interactions~\cite{Wang19NGCF,lee2018goccf,He20LightGCN,mao2021simplex,Mao21UltraGCN,choi2021ltocf,Shen21GFCF,hu2022mgdcf,chen20LRGCCF,kong2022hmlet,hong2022timekit,fan2022GTN,liu2021IMP-GCN}. However, the inherent data sparsity in recommendation scenarios, where most users interact with only a few items, pose a challenge. CL addresses this by exposing the model to more diverse training environments, thereby stabilizing the training process and mitigating overfitting.

\begin{figure}[t!]
    \vspace{1em}
    \centering
    \includegraphics[width=0.7\columnwidth]{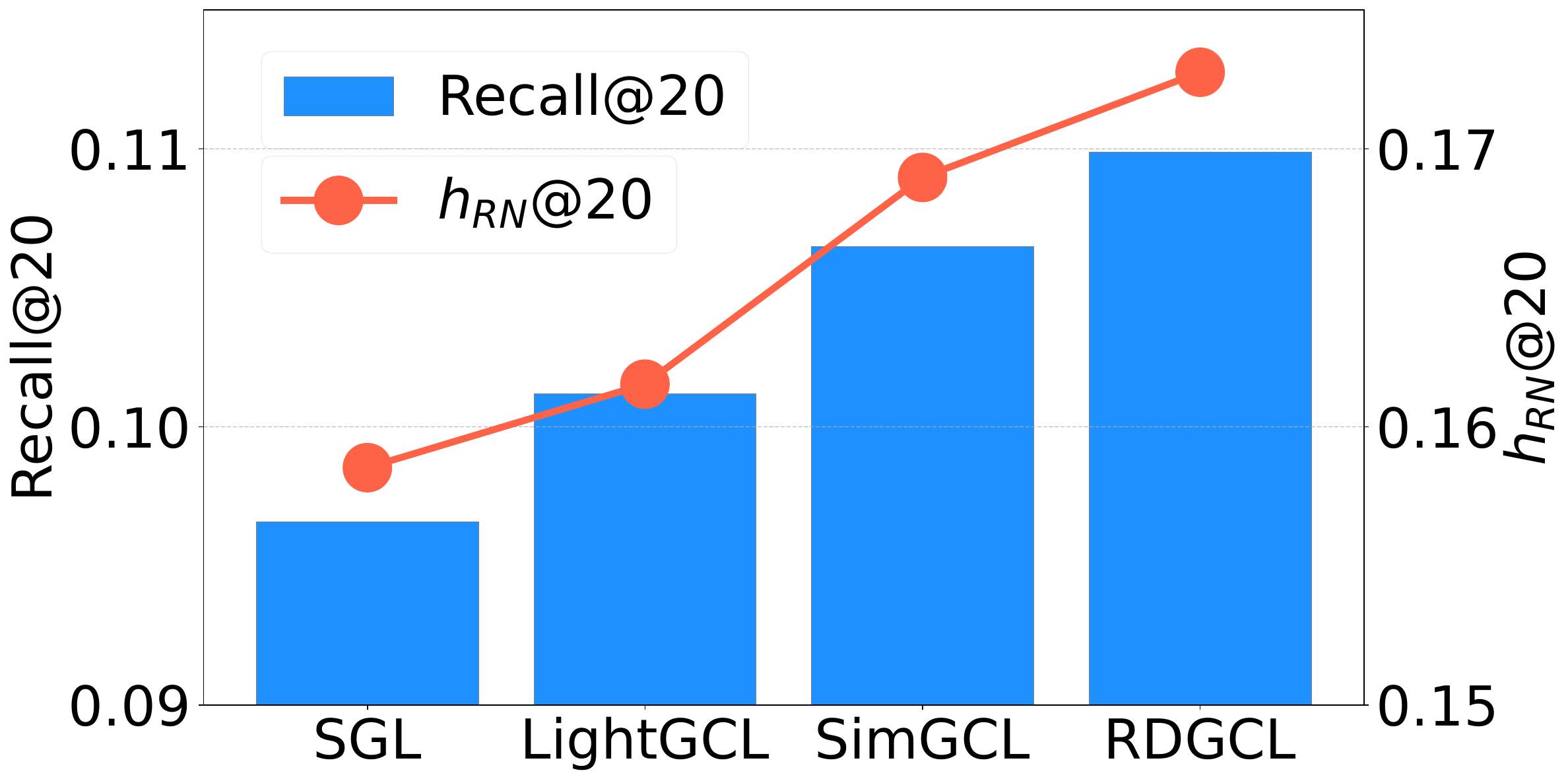}
    \caption{Comparison in terms of Recall@20 and $h_{\text{RN}}$@20, the harmonic mean of the recall and novelty (see Sec.~\ref{sec:trade}), on Yelp. Our RDGCL outperforms SGL~\cite{Wu2021SGL}, SimGCL~\cite{xu2023simdcl}, and LightGCL~\cite{cai2023lightgcl}.}
    \label{fig:coverage}
\end{figure}

\begin{table}[t]
    \small
    \centering
    \setlength{\tabcolsep}{0.8pt}
    \begin{tabular}{l lcc}\toprule
        Model & View generation        & Low-pass filter    & High-pass filter\\\midrule
        SGL & Node/edge dropout   & \cmark & \xmark \\
        SimGCL & Injecting noises into embs. & \cmark & \xmark \\
        LightGCL & Filtering graphs with SVD  & \cmark & \xmark \\\midrule
        \textbf{RDGCL} & Low vs. High-pass embs. & Diffusion eq. & Reaction eq. \\
        \bottomrule
    \end{tabular}
    \caption{Comparison of existing CL-based methods that differ at three points: i) how to generate views for CF, ii) using low-pass filters, and iii) using high-pass filters.}
    \label{tab:comp}
\end{table}

Existing CL-based recommendation models rely on graph augmentation techniques to generate contrastive views. These techniques typically involve perturbing graph structures or adding noise to node embeddings. As in Table~\ref{tab:comp}, SGL~\cite{Wu2021SGL} perturbs graph structures, SimGCL~\cite{yu2022SimGCL} injects uniform noise into embeddings, and LightGCL~\cite{cai2023lightgcl} reconstructs graph structures through singular value decomposition (SVD).
However, the current paradigm of CL-based CF has several limitations: 
\begin{enumerate}
    \item Graph augmentations can introduce noise and redundancies, potentially degrading the quality of the learned node representation. 
    \item These augmentations may not generate sufficient diversity and contrast in node representations (see Fig.~\ref{fig:coverage}). 
    \item Existing methods using LightGCN use low-pass graph filters (see Table 1), overlooking the importance of high-pass filters. 
\end{enumerate}
Recent research has highlighted the limitations of using only low-frequency signals (i.e., popular items)~\cite{Peng2022Less,choi2023bspm,du2023fearec} in recommender systems. Models based on low-pass filters, such as LightGCN struggle to recommend novel or diverse items that reflect unique user preferenes.

To address these limitations, the key design points in our proposed method are twofold: i) a new GCN-based network is proposed for CF, and ii) a new CL method for it is designed. Our method with the two contributions is called \emph{reaction-diffusion graph contrastive learning} (RDGCL) since those design points are greatly inspired by the reaction-diffusion equation.

Our approach uses the reaction-diffusion equation in the context of GCNs, where the diffusion equation makes the embeddings of neighboring nodes similar~\cite{wang2021dgc}, and the reaction equation makes them dissimilar (high-pass filter)~\cite{choi2023gread,wang2023acmp}. In the perspective of graph signal processing, the diffusion (resp. reaction) equation corresponds to the low-pass (resp. high-pass) filter.

RDGCL differs from other CL-based CF methods by using \emph{a single pass} architecture. For instance, LightGCL has two GCN instances, one for the main CF task and the other for the augmented graph view purpose, which we call \emph{two passes}. Since RDGCL has both diffusion and reaction layers internally, this allows for efficient CL training between these layers without the need for separate graph augmentation steps (see Fig.~\ref{fig:overview}).

This paper presents a comprehensive evaluation of RDGCL using 5 benchmark datasets and 14 baselines. Our experimental results demonstrate the superiority of RDGCL in terms of recommendation accuracy, coverage, and novelty. 
The main contributions of this paper are:
\begin{itemize}
    \item We propose RDGCL, a novel CL-based CF approach incorportaing both diffusion (low-pass filtering) and reaction (high-pass filtering) equations in its neural network design and CL training method.
    \item To our knowledge, RDGCL is the first to apply the reaction-diffusion equation for CL-based CF methods.
    \item RDGCL outperforms 14 baselines on 5 benchmark datasets. 
    \item In terms of relevance and diversity metrics (e.g. coverage and novelty), RDGCL achieves the most balanced performance, recommending more diverse and novel items while maintaining high accuracy (see Fig.~\ref{fig:coverage}).
\end{itemize}

\section{Preliminaries \& Related Work}\label{sec:related}
\subsection{Graph Filters and GCN-based CFs}
Let $\mathbf{R} \in \{0,1\}^{|\mathcal{U}| \times |\mathcal{V}|}$ be an interaction matrix, where $\mathcal{U}$ and $\mathcal{V}$ is sets of users and items. $\mathbf{R}_{u,v}=1$ \textit{iff} an interaction $(u,v)$ is observed and 0 otherwise.The adjacency matrix $\mathbf{A} \in \mathbb{R}^{N \times N}$, where $N=|\mathcal{U}|+|\mathcal{V}|$, represent the graph structure. The Laplacian matrix is defined as $\mathbf{L}=\mathbf{D} - \mathbf{A}\in \mathbb{R}^{N \times N}$, with $\mathbf{D}$ being the diagonal degree matrix. The symmetric normalized adjacency and Laplacian matrices are defined as $\tilde{\mathbf{A}} = \bar{\mathbf{D}}^{-\frac{1}{2}}\bar{\mathbf{A}}\bar{\mathbf{D}}^{-\frac{1}{2}}$, and $\tilde{\mathbf{L}} = \mathbf{I} - \bar{\mathbf{A}}$, 
where $\bar{\mathbf{D}}=\mathbf{D}+\mathbf{I}$ and $\bar{\mathbf{A}}=\mathbf{A}+\mathbf{I}$.

The operation $\tilde{\mathbf{L}}\mathbf{x}$ can be interpreted as a filter that modifies the frequency components of the graph signal $\mathbf{x}$~\cite{chung1997spectral}. The Laplacian filter enhances signal components associated with higher eigenvalues ${\gamma}_i \in (1,2)$ while reducing those with lower eigenvalues ${\gamma}_i \in [0,1]$. 
This behavior characterizes Laplacian matrices as high-pass filters, emphasizingdifferences in node features~\cite{Ekambaram2013lens}. In contrast, normalized adjacency matrices function as low-pass filters~\cite{nt2019revisiting}, reducing non-smooth signal components. This is due to all eigenvalues of the adjacency matrices being less than 1, i.e. ${\gamma}_i \in (-1, 1]$.

Most GCN-based CF approaches, including LightGCN~\cite{He20LightGCN}, use low-pass filters to enhance the smoothness of node representations.
While some recent studies use high-pass filters~\cite{Peng2022Less,choi2023bspm,shin2023attentive}, their application in generating views for CL in recommendation remains unexplored. Our key idea is to apply high-pass graph filters in the CL framework for recommendation.

\subsection{Contrastive Learning for Recommendation}
Deep learning-based recommender systems have shown remarkable performance in recent years. However, they suffer from data sparsity and cold start problems, since they rely heavily on labels, i.e., positive user-item interactions~\cite{yu2018ifbpr,you2020GraphCL}. To address these problems, self-supervised methods, particulary CL-based CF methods~\cite{jing2023survey,Wu2021SGL} have emerged as promising outcomes.

SGL~\cite{Wu2021SGL} first applied the CL to graph-based recommendation, using LightGCN~\cite{He20LightGCN} as its graph encoder. It introduces three operators to generate augmented views: node dropouts, edge dropouts, and random walks. By contrasting these augmented views, it improves the recommendation accuracy, especially for long-tail items, and the robustness against interaction noises. For CL, InfoNCE~\cite{oord2018infonce} is defined as follows:
\begin{align}
\mathcal{L}_{CL} = \sum_{i \in \mathcal{B}} -log\frac{\exp(\text{sim}(\mathbf{e}_i',\mathbf{e}_i'')/\tau)}{\sum_{j \in \mathcal{B}}\exp(\text{sim}(\mathbf{e}_i'\mathbf{e}_j'')/\tau)},\label{eq:general_contrastive_loss}
\end{align} where $i$, $j$ are a user and an item in a mini-batch $\mathcal{B}$ respectively, $\text{sim}(\cdot)$ is the cosine similarity, $\tau$ is the temperature, and $\mathbf{e}'$, $\mathbf{e}''$ are augmented node representations. The CL loss increases the alignment between  $\mathbf{e}_i'$ and $\mathbf{e}_i''$ nodes, viewing the representations of the same node $i$ as positive pairs. Simultaneously, it minimizes the alignment between the node representations of $\mathbf{e}_i'$ and $\mathbf{e}_j''$, viewing the representations of the different nodes $i$ and $j$ as negative pairs.

\begin{figure}[t]
    \centering
    \subfigure[SimGCL]{\includegraphics[width=0.49\columnwidth]{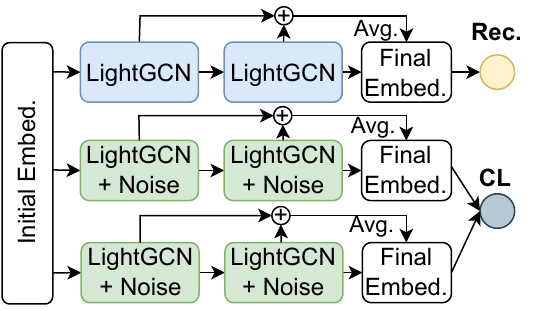}}
    \subfigure[LightGCL]{\includegraphics[width=0.49\columnwidth]{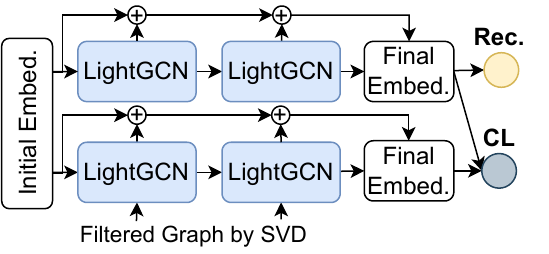}}
    \subfigure[XSimGCL]{\includegraphics[width=0.49\columnwidth]{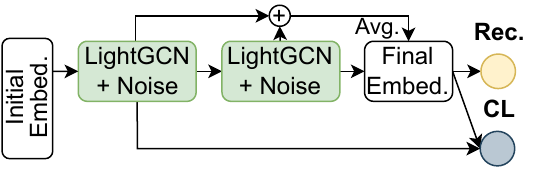}}
    \subfigure[RDGCL]{\includegraphics[width=0.49\columnwidth]{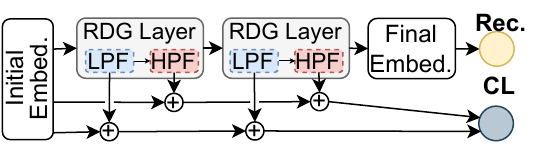}}
    \caption{The architectures of SimGCL, LightGCL, XSimGCL, and RDGCL.}
    \label{fig:scheme}
\end{figure}

SimGCL~\cite{yu2022SimGCL} simplifies the graph augmentation process for its CL by perturbing node representations with random noise. XSimGCL~\cite{yu2022xsimgcl} replaces the final-layer CL of SimGCL with a cross-layer CL approach --- our RDGCL also follows this cross-layer CL approach. It only uses one GCN-based encoder and contrasts the embeddings of different layers, and this cross-layer CL reduces the computational complexity since it has only one neural network. 
NCL~\cite{lin2022ncl} does not change the existing graph structure to create contrastive views and finds the neighbors of a user or item in semantic space to suggest new contrastive learning. NCL incorporates semantic neighbors into the prototype-contrastive objective to find potential neighbor relationships in semantic space.
LightGCL~\cite{cai2023lightgcl} proposes a SVD-based graph augmentation strategy to effectively distill global collaborative signals. In specific, SVD is first performed on the adjacency matrix. Then, the list of singular values is truncated to retain the largest values, i.e., the ideal low-pass filter, and then its truncated matrix is used to purify the adjacency matrix.
As shown in Fig.~\ref{fig:scheme}, existing CL-based recommender systems are limited to low-pass filters since i) their backbones are mostly LightGCN and ii) they augment views with low-pass filters.

\subsection{Reaction-Diffusion Equations}
Reaction-diffusion equations are partial differential equations that describe how the concentration of substances distributed in space changes under the influence of local chemical reactions and diffusion~\cite{alan1952morphogenesis,kondo2010reaction}. A general form of a reaction-diffusion equation is:
\begin{align}
\frac{\partial u}{\partial t}=\nabla^{2}u+R(u),
\end{align}
where $u(x,t)$ is the concentration of a substance at position $x$ and time $t$, $\nabla^{2}$ is the Laplace operator, and $R(u)$ is the reaction term. 
There are different types of reaction terms to describe pattern formation phenomena in various biological~\cite{fisher1937wave} and chemical~\cite{allen1979microscopic,pinar2021analytical} systems, and image processing~\cite{turk1991generating,witkin1991reaction,plonka2008nonlinear,chen2016trainable}.
On graphs, reaction-diffusion equations can be discretized using finite difference methods. Using the Euler scheme, we can approximate the reaction-diffusion equation in the context of graph signal processing as:
\begin{align}
u(t+\Delta t)=u(t)+\Delta t(-\tilde{\mathbf{L}}u(t)+R(u(t))),
\end{align}
where $\Delta t$ is the time step size and $u(t) \in \mathbb{R}^{N}$ is the graph signal at time $t$. This equation updates the graph signal by applying diffusion and reaction terms at each time step. 

Inspired by reaction-diffusion filters in image processing that iterate between blurring (low-pass) and sharpening (high-pass) processes~\cite{plonka2008nonlinear}, we use graph high-pass filters to design the interaction of nodes in the reaction term.

\subsection{Neural Ordinary Differential Equations}
Neural ordinary differential equations (NODEs)~\cite{chen2018NODE} solve the initial value problem to calculate $\mathbf{h}(t_{i+1})$ from $\mathbf{h}(t_i)$:
\begin{align}\label{eq:node}
\mathbf{h}(t_{i+1}) = \mathbf{h}(t_i) + \int_{t_i}^{t_{i+1}} f(\mathbf{h}(t_i), t;\mathbf{\theta}_f) dt,
\end{align}
where $f(\cdot)$, parameterized by $\mathbf{\theta}_f$, approximates the time-derivative of $\mathbf{h}$. Various ODE solvers can be used, with the Euler method being a simple example:
\begin{align}\label{eq:euler}
\mathbf{h}(t + s) = \mathbf{h}(t) + s \cdot f(\mathbf{h}(t)),
\end{align}
where $s$ is the step size. Eq.~\eqref{eq:euler} is identical to a residual connection when $s=1$ and therefore NODEs are a continuous generalization of residual networks.

Continuous GCNs~\cite{xhonneux2019CGNN,poli2019gde} use NODEs to transform discrete GCN layers into continuous layers, which can be interpreted in terms of diffusion processes~\cite{wang2021dgc,chamberlain2021grand}. 
Since our goal is to design our method to bring the reaction-diffusion process into a GCN and CL-based recommender system, we design our method based on NODEs.

\section{Proposed Method}\label{sec:method}

\subsection{Overall Architecture}
\begin{figure}[t]
    \centering
    \includegraphics[width=\columnwidth]{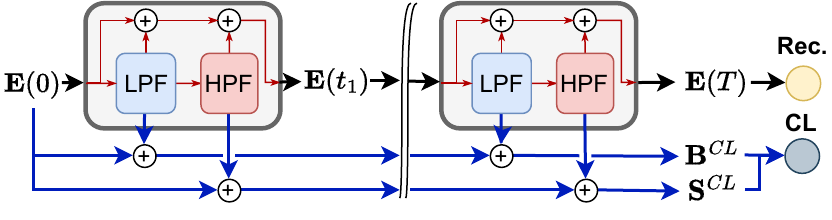}
    \caption{The illustration of our RDGCL where we solve our reaction-diffusion system with the Euler method.
    LPF (resp. HPF) stands for the low (resp. high)-pass filter.}
    \label{fig:overview}
\end{figure}

Our RDGCL architecture, shown in Fig.~\ref{fig:overview} begins with an initial embedding $\mathbf{E}(0)$ that evolves over time $t \in [0,T]$ through a reaction-diffusion graph (RDG) layer. This embedding evolutionary process is described by:
\begin{align}\label{eq:rdnode}
    \mathbf{E}(T) &= \mathbf{E}(0) + \int_{0}^{T}f(\mathbf{E}(t))dt,
\end{align}
where $\mathbf{E}(t) \in \mathbb{R}^{N \times D}$ is the node embedding matrix at time $t$ with $D$ dimensions. $f(\mathbf{E}(t))$ is a RDG layer which outputs $\frac{d \mathbf{E}(t)}{dt}$.
The iterative process of the RDG layer consists of:
\begin{enumerate}
    \item The RDG layer applies a low-pass filter (e.g., a diffusion process) to $\mathbf{E}(t_i)$ to derive its low-pass filtered embedding $\mathbf{B}(t_i)$.
    \item It then applies a high-pass filter (e.g., a reaction process) to $\mathbf{B}(t_i)$ to derive $\mathbf{E}(t_{i+1})$. 
\end{enumerate}

RDGCL uses $\mathbf{E}(T)$ for recommendation while contrasting two views, $\mathbf{B}^{CL}$ and $\mathbf{S}^{CL}$, representing low-pass and high-pass information respectively.  This approach is called \emph{cross-layer CL} or \emph{single-pass CL}.

\subsection{Reaction-Diffusion Graph (RDG) Layer}
Our RDG layer is defined by:
\begin{align}\label{eq:rd}
f(\mathbf{E}(t)) &:= \frac{d \mathbf{E}(t)}{dt} = -\tilde{\mathbf{L}}\mathbf{E}(t) + \alpha R(\mathbf{E}(t)),
\end{align}
where $R(\cdot)$ is the reaction term, and $\alpha$ is the reaction rate coefficient to (de-)emphasize the reaction term. 

\subsubsection{Diffusion Process as Low-pass Filter}
Multiplying with the adjacency matrix ($\mathbf{A}$ or $\tilde{\mathbf{A}}$) is considered a low-pass filtering operation. Many GCNs can be generalized to diffusion process~\cite{wang2021dgc,chamberlain2021grand}:
\begin{align}
\mathbf{B}(t) = \mathbf{E}(t)-\tilde{\mathbf{L}}\mathbf{E}(t)=\mathbf{E}(t) + (\tilde{\mathbf{A}}-\mathbf{I})\mathbf{E}(t) = \tilde{\mathbf{A}}\mathbf{E}(t).\label{eq:diffusion}
\end{align}

\subsubsection{Reaction Process as High-pass Filter}
The reaction process acting as a high-pass filter, is applied to $\mathbf{B}(t)$:
\begin{align}
R(\mathbf{E}(t)) &:= \tilde{\mathbf{L}}\mathbf{B}(t)=\tilde{\mathbf{L}}\tilde{\mathbf{A}}\mathbf{E}(t).\label{eq:reaction}
\end{align}
The reaction process counteracts the oversmoothing effect in GCNs. Oversmoothing tends to standardize user preferences towards popular items, worsening the long-tail problem~\cite{zhao2022investigating,zhou2023adaptive}. RDGCL emphasizes the distinctions among nodes by including high-pass filtering, thereby recommending long-tail items to users. In Sec.~\ref{sec:trade} and Sec.~\ref{sec:robustness}, we will provide empirical evidence of the high-pass filter's effectiveness in promoting diversity and alleviating the popularity bias.

\subsection{Cross-layer Contrastive Learning}
We extract low-pass (or diffusion) and high-pass (or reaction) views from Eq.~\eqref{eq:rd} along $[0,T]$:
\begin{align}\label{eq:views}
\mathbf{B}^{CL} = \mathbf{E}(0)+\sum_{i=1}^K\mathbf{B}(t_i),\quad \mathbf{S}^{CL} = \mathbf{E}(0)+\sum_{i=1}^K R(\mathbf{E}(t_i)),
\end{align}
where we solve Eq.~\eqref{eq:rdnode} via $K$ steps with an ODE solver. 

We perform the CL training by directly contrasting two views of reaction and diffusion using InfoNCE loss: 
\begin{align}\label{eq:contrastive_loss}
\mathcal{L}_{CL} = \sum_{i \in \mathcal{B}} -log\frac{\exp(\text{sim}(\mathbf{b}^{CL}_{i},\mathbf{s}^{CL}_{i})/\tau)}{\sum_{j \in \mathcal{B}}\exp(\text{sim}(\mathbf{b}^{CL}_{i},\mathbf{s}^{CL}_{j})/\tau)},
\end{align}where $\mathbf{b}^{CL}_i$,  $\mathbf{s}^{CL}_{i}$, and $\mathbf{s}^{CL}_{j}$ are node representations from Eq.~\eqref{eq:views}.
The final objective function is as follows:
\begin{align}\label{eq:obejective_function}
\mathcal{L} =\mathcal{L}_{BPR}+\lambda_1 \cdot \mathcal{L}_{CL}+\lambda_2 \cdot \Vert\mathbf{\Theta}\Vert^2_2,
\end{align}where $\mathcal{L}_{BPR}$ is the Bayesian personalized ranking (BPR) loss and $\mathbf{\Theta} = \mathbf{E}(0)$.

\subsection{Model Complexity}
We analyze the time complexity of RDGCL compared to LightGCN, SGL, and SimGCL. Let $|\mathbf{A}|$ be the number of edges, $|\mathcal{B}|$ the batch size, $M$ the number of nodes in a batch, $K$ the number of layers or ODE steps, and $\rho$ the edge keeping-rate in SGL. Table~\ref{tab:time} summarizes the time complexity which gives the following findings:
\begin{itemize}
    \item RDGCL, LightGCN and SimGCL, requires $\mathcal{O}(2|\mathbf{A}|)$ to construct the adjacency matrix, while SGL needs nearly triple this cost due to graph augmentation.
    \item The computational cost per step of RDGCL is lower than that of SGL, requiring only two matrix multiplication operations in Eq.~\eqref{eq:rd}. 
    \item For the CL loss computation, the complexity of RDGCL is the same as those of other models, which is $\mathcal{O}(|\mathcal{B}|D + |\mathcal{B}|MD)$, where $\mathcal{O}(|\mathcal{B}|D)$ and $\mathcal{O}(|\mathcal{B}|MD)$ are for positive and negative views, respectively. For brevity, we show it as $\mathcal{O}(|\mathcal{B}|MD)$.
\end{itemize}
This analysis shows the computational efficiency of RDGCL relative to existing methods.

\begin{table}[t]
    \small
    \setlength{\tabcolsep}{1pt}
    \centering
    
    \resizebox{0.99\linewidth}{!}{
    \begin{tabular}{c cccc}\toprule
        Component        & LightGCN      & SGL           & SimGCL           & RDGCL\\\midrule
        Adj. Matrix & $\mathcal{O}(2|\mathbf{A}|)$ & $\mathcal{O}((2+4\rho)|\mathbf{A}|)$ & $\mathcal{O}(2|\mathbf{A}|)$ & $\mathcal{O}(2|\mathbf{A}|)$\\
        GCN   & $\mathcal{O}(2|\mathbf{A}|KD)$ & $\mathcal{O}((2+4\rho)|\mathbf{A}|KD)$ & $\mathcal{O}(6|\mathbf{A}|KD)$ & $\mathcal{O}(4|\mathbf{A}|KD)$\\
        CL       & N/A & $\mathcal{O}(|\mathcal{B}|MD)$ & $\mathcal{O}(|\mathcal{B}|MD)$ & $\mathcal{O}(|\mathcal{B}|MD)$ \\ \bottomrule
    \end{tabular}
    }
    \caption{The comparison of analytical time complexity}
    \label{tab:time}
\end{table}

\begin{table*}[t]
    \small
    \setlength{\tabcolsep}{1pt}
    \centering
    \resizebox{0.99\linewidth}{!}{
    \begin{tabular}{cc ccccccccccccccc cc}\toprule
    \multicolumn{1}{c}{Data} &
      \multicolumn{1}{c}{Metric} &
      \multicolumn{1}{c}{LightGCN} &
      \multicolumn{1}{c}{LT-OCF} &
      \multicolumn{1}{c}{HMLET} &
      \multicolumn{1}{c}{SGL} &
      \multicolumn{1}{c}{SimGRACE} &
      \multicolumn{1}{c}{GCA} &
      \multicolumn{1}{c}{HCCF} &
      \multicolumn{1}{c}{SHT} &
      \multicolumn{1}{c}{SimGCL} &
      \multicolumn{1}{c}{XSimGCL} &
      \multicolumn{1}{c}{NCL} &
      \multicolumn{1}{c}{LightGCL} &
      \multicolumn{1}{c}{GF-CF} &
      \multicolumn{1}{c}{BSPM} &
      \multicolumn{1}{c}{\textbf{RDGCL}} & \textit{Imp.} \\ \midrule
    \multirow{4}{*}{\rotatebox[origin=c]{90}{Yelp}}
            & Recall@20 & 0.0826 & 0.0947 & 0.0859 & 0.0967 & 0.0899 & 0.0779 & 0.0995 & 0.0853 & \SECOND{0.1065} & 0.0974 & 0.0956 & 0.1012 & 0.1043 & \THIRD{0.1059} & \BEST{0.1099} & \textit{3.19\%}\\
            & NDCG@20   & 0.0690 & 0.0800 & 0.0699 & 0.0824 & 0.0775 & 0.0670 & 0.0842 & 0.0719 & \THIRD{0.0912} & 0.0823 & 0.0812 & 0.0870 & 0.0890 & \SECOND{0.0913} & \BEST{0.0939} & \textit{2.77\%}\\
            & Recall@40 & 0.1346 & 0.1531 & 0.1388 & 0.1544 & 0.1443 & 0.1279 & 0.1578 & 0.1382 & \SECOND{0.1688} & 0.1564 & 0.1522 & 0.1591 & 0.1659 & \THIRD{0.1677} & \BEST{0.1721} & \textit{1.95\%}\\
            & NDCG@40   & 0.0882 & 0.1016 & 0.0898 & 0.1032 & 0.1032 & 0.0851 & 0.1056 & 0.0913 & 0.1038 & 0.1040 & 0.1019 & 0.1081 & \THIRD{0.1115} & \SECOND{0.1139} & \BEST{0.1165} & \textit{2.28\%}\\ \midrule
    \multirow{4}{*}{\rotatebox[origin=c]{90}{Gowalla}}
            & Recall@20 & 0.1294 & 0.2215 & 0.2157 & \THIRD{0.2410} & 0.1519 & 0.1899 & 0.2222 & 0.1877 & 0.2382 & 0.2314 & 0.1933 & 0.2351 & 0.2347 & \SECOND{0.2455} & \BEST{0.2564} & \textit{4.43\%}\\
            & NDCG@20   & 0.0781 & 0.1287 & 0.1270 & \THIRD{0.1427} & 0.0850 & 0.1100 & 0.1298 & 0.1119 & 0.1412 & 0.1380 & 0.1129 & 0.1386 & 0.1382 & \SECOND{0.1472} & \BEST{0.1549} & \textit{5.19\%}\\
            & Recall@40 & 0.1869 & 0.3131 & 0.3010 & \THIRD{0.3313} & 0.2225 & 0.2690 & 0.3106 & 0.2671 & 0.3265 & 0.3224 & 0.2773 & 0.3251 & 0.3266 & \SECOND{0.3342} & \BEST{0.3460} & \textit{3.53\%}\\
            & NDCG@40   & 0.0932 & 0.1529 & 0.1494 & \THIRD{0.1665} & 0.1034 & 0.1309 & 0.1528 & 0.1325 & 0.1644 & 0.1619 & 0.1347 & 0.1622 & 0.1624 & \SECOND{0.1707} & \BEST{0.1783} & \textit{4.45\%}\\ \midrule 
    \multirow{4}{*}{\rotatebox[origin=c]{90}{\parbox{1.2cm}{Amazon-\\Electronics}}}
            & Recall@20 & 0.1342 & 0.1319 & 0.1355 & \SECOND{0.1393} & 0.1360 & 0.1254 & 0.0597 & 0.1255 & 0.1371 & 0.1295 & \THIRD{0.1377} & 0.1306 & 0.1306 & 0.1311 & \BEST{0.1407} & \textit{1.22\%} \\
            & NDCG@20   & 0.0783 & 0.0792 & 0.0783 & \THIRD{0.0797} & 0.0789 & 0.0727 & 0.0338 & 0.0742 & 0.0777 & 0.0748 & \SECOND{0.0803} & 0.0771 & 0.0771 & 0.0792 & \BEST{0.0812}& \textit{1.12\%} \\
            & Recall@40 & 0.1873 & 0.1850 & 0.1955 & 0.1916 & \THIRD{0.1963} & 0.1306 & 0.0918 & 0.1816 & 0.1939 & 0.1833 & \SECOND{0.1978} & 0.1908 & 0.1907 & 0.1742 & \BEST{0.2015} & \textit{1.87\%} \\
            & NDCG@40   & 0.0933 & \THIRD{0.0958} & 0.0948 & 0.0944 & 0.0954 & 0.0771 & 0.0426 & 0.0895 & 0.0934 & 0.0897 & \SECOND{0.0967} & 0.0934 & 0.0934 & 0.0912 & \BEST{0.0977} & \textit{1.03\%} \\ \midrule
    \multirow{4}{*}{\rotatebox[origin=c]{90}{\parbox{1.2cm}{Amazon-\\CDs}}}
            & Recall@20 & 0.1111 & 0.1572 & 0.1532 & \THIRD{0.1778} & 0.1160 & 0.1510 & 0.0658 & 0.1309 & 0.1722 & 0.1430 & 0.1667 & \SECOND{0.1804} & 0.1394 & 0.1443 & \BEST{0.1823} & \textit{1.05\%}\\ 
            & NDCG@20   & 0.0659 & 0.0981 & 0.0953 & \THIRD{0.1163} & 0.0720 & 0.1000 & 0.0434 & 0.0815 & 0.1132 & 0.0862 & 0.1046 & \SECOND{0.1181} & 0.0912 & 0.0929 & \BEST{0.1202} & \textit{1.69\%}\\
            & Recall@40 & 0.1695 & 0.2196 & 0.2141 & \THIRD{0.2401} & 0.1678 & 0.2057 & 0.1076 & 0.1890 & 0.2319 & 0.2082 & 0.2333 & \SECOND{0.2418} & 0.1940 & 0.1974 & \BEST{0.2449} & \textit{1.28\%}\\
            & NDCG@40   & 0.0821 & 0.1157 & 0.1124 & \THIRD{0.1338} & 0.0886 & 0.1154 & 0.0542 & 0.0978 & 0.1230 & 0.1044 & 0.1231 & \SECOND{0.1353} & 0.1066 & 0.1085 & \BEST{0.1377} & \textit{1.77\%}\\ \midrule
    \multirow{4}{*}{\rotatebox[origin=c]{90}{Tmall}}
            & Recall@20 & 0.0717 & 0.0737 & 0.0676 & \SECOND{0.1035} & 0.0705 & 0.0730 & 0.0930 & 0.0706 & 0.0993 & 0.0880 & 0.0684 & \THIRD{0.1033} & 0.0879 & 0.0879 & \BEST{0.1040} & \textit{0.44\%}\\
            & NDCG@20   & 0.0498 & 0.0515 & 0.0464 & \SECOND{0.0745} & 0.0488 & 0.0511 & 0.0663 & 0.0492 & \THIRD{0.0712} & 0.0626 & 0.0470 & 0.0626 & 0.0612 & 0.0612 & \BEST{0.0753} & \textit{0.64\%}\\
            & Recall@40 & 0.1126 & 0.1168 & 0.1059 & \SECOND{0.1545} & 0.1104 & 0.1124 & 0.1427 & 0.1118 & \THIRD{0.1505} & 0.1370 & 0.1082 & 0.1370 & 0.1379 & 0.1380 & \BEST{0.1559} & \textit{0.86\%}\\
            & NDCG@40   & 0.0640 & 0.0665 & 0.0598 & \SECOND{0.0922} & 0.0626 & 0.0648 & 0.0834 & 0.0634 & \THIRD{0.0890} & 0.0795 & 0.0607 & 0.0795 & 0.0784 & 0.0785 & \BEST{0.0933} & \textit{0.85\%}\\ \bottomrule
    \end{tabular}
    }
    \caption{Performance comparison to popular graph-based CF and CL models. We highlight the best 3 results in \BEST{red} (first), \SECOND{blue} (second), and \THIRD{purple} (third). \textit{Imp.} stands for relative improvement over second-best performance.
    See \emph{Appendix~\ref{app:cohen}} for Cohen's effect size~\cite{sullivan2012using} results for statistical testing.}
    \label{tab:result}
\end{table*}


\subsection{Property of RDG Layer}
Under Euler discretization with $dt=1$, $T=1$, and $\alpha=1$, Eq.~\eqref{eq:rdnode} can be interpreted as applying a $2\tilde{\mathbf{A}}-\tilde{\mathbf{A}}^2$ filter to $\mathbf{E}(t)$. The derivation of this filter is provided in \emph{Appendix~\ref{app:rdg_derivation}}. 

\begin{theorem}\label{thrm:filter}
The RDG layer applies a $2\tilde{\mathbf{A}} - \tilde{\mathbf{A}}^2$ filter, which emphasizes mid-to-high frequency components of the graph signal more than the $\tilde{\mathbf{A}}$ filter.
\end{theorem}

This filter emphasizes higher-frequency signals while attenuating low-frequency ones~\cite{singer2009diffusion}, unlike the $\tilde{\mathbf{A}}$ filter. This property captures unique user preferences, potentially increasing recommendation diversity (see Sec.\ref{sec:trade}). The proof is in \emph{Appendix~\ref{app:proof}}.

\subsection{Relation to Other Methods}
RDGCL uses a \emph{single pass} for CL and CF tasks, in contrast to SimGCL and LightGCL, which use separate passes.
This approach improves efficiency and consistency in representation learning. While XSimGCL shares similarities with RDGCL in using a single-pass, cross-layer CL method, it remains limited to low-pass filters. RDGCL adopts the reaction-diffusion system to contrast the information from the different graph signals. However, SimGCL and LightGCL perform the CL training for their final embeddings only, overlooking their intermediate embeddings. This enables RDGCL to capture more diverse and complementary features from the graph structure.

RDGCL can be viewed as an extension of LT-OCF, a continuous-time generalization of LightGCN. RDGCL reduces to LT-OCF if the reaction term and CL are removed. In contrast, BSPM applies low-pass and high-pass filters to the interaction matrix sequentially and only once, while RDGCL integrates these processes iteratively as per Eq~\eqref{eq:rd}.

\section{Experiments}\label{sec:exp}

\subsection{Experimental Environments}
\subsubsection{Datasets and Baselines}
We evaluate our model and baselines with 5 real-world benchmark datasets: Yelp, Gowalla, Amazon-Electronics, Amazon-CDs, and Tmall~\cite{Wang19NGCF,He20LightGCN, harper2015movielens}.
The detailed dataset statistics are presented in \emph{Appendix~\ref{app:data}}.
We compare our model with the 14 baselines with 4 categories:
\begin{itemize}
    \item Graph-based CFs include LightGCN~\cite{He20LightGCN}, LT-OCF~\cite{choi2021ltocf}, HMLET~\cite{kong2022hmlet}, GF-CF~\cite{Shen21GFCF}, and BSPM~\cite{choi2023bspm};
    \item Graph CL methods for other tasks include SimGRACE~\cite{xia2022SimGRACE} and GCA~\cite{zhu2021GCA};
    \item Hyepergraph-based CFs include HCCF~\cite{xia2022HCCF} and SHT~\cite{xia2022SHT};
    \item Graph CL methods for CF include SGL~\cite{Wu2021SGL}, SimGCL~\cite{yu2022SimGCL}, XSimGCL~\cite{yu2022xsimgcl}, NCL~\cite{lin2022ncl}, and LightGCL~\cite{cai2023lightgcl}.
\end{itemize}

\subsubsection{Evaluation Protocols and Hyperparameters}
We strictly follow evaluation protocols of \citet{cai2023lightgcl}, using Recall@20/40 and NDCG@20/40.
We use hyperparameters for all baseline models based on their recommended ranges. The detailed search ranges and the best settings for RDGCL on each dataset are provided in \emph{Appendix~\ref{app:exp_detail}}.

\subsection{Experimental Results}
Table~\ref{tab:result} shows Recall@20/40 and NDCG@20/40. Our RDGCL clearly shows the highest accuracy in all cases. Specifically for Gowalla, RDGCL's NDCG@20 is 5.19\% higher than the best baseline.
SGL and SimGCL perform well in some cases, with only SimGCL comparable to the accuracy of RDGCL on Yelp. SGL comparable to RDGCL on Tmall. However, the accuracy gap between RDGCL and SGL is still non-trivial. XSimGCL, known to be efficient and perform well, does not perform well on datasets we use. 

To further verify the performance of RDGCL, we compare it with two other recently proposed methods. GF-CF and BSPM show good performance, occupying the 2nd and 3rd places for Yelp and Gowalla. 
However, no existing methods are comparable to our RDGCL for all datasets. 

\subsection{Trade-off Among Recall, Coverage, and Novelty} \label{sec:trade}
We analyze the balanced nature of our model in terms of recall, coverage, and novelty. To provide a comprehensive evaluation, we use the harmonic means of recall with both coverage and novelty. The detailed formulation of these harmonic means is provided in \emph{Appendix~\ref{app:harmonic_mean}}.
Coverage~\cite{Herlocker2004Coverage} refers to the range of items that models can recommend, and novelty~\cite{zhou2010solving} measures how unexpected the recommended items are compared to their global popularity. 

\begin{figure}[t!]
    \centering
    \subfigure[Yelp]{\includegraphics[width=0.49\columnwidth]{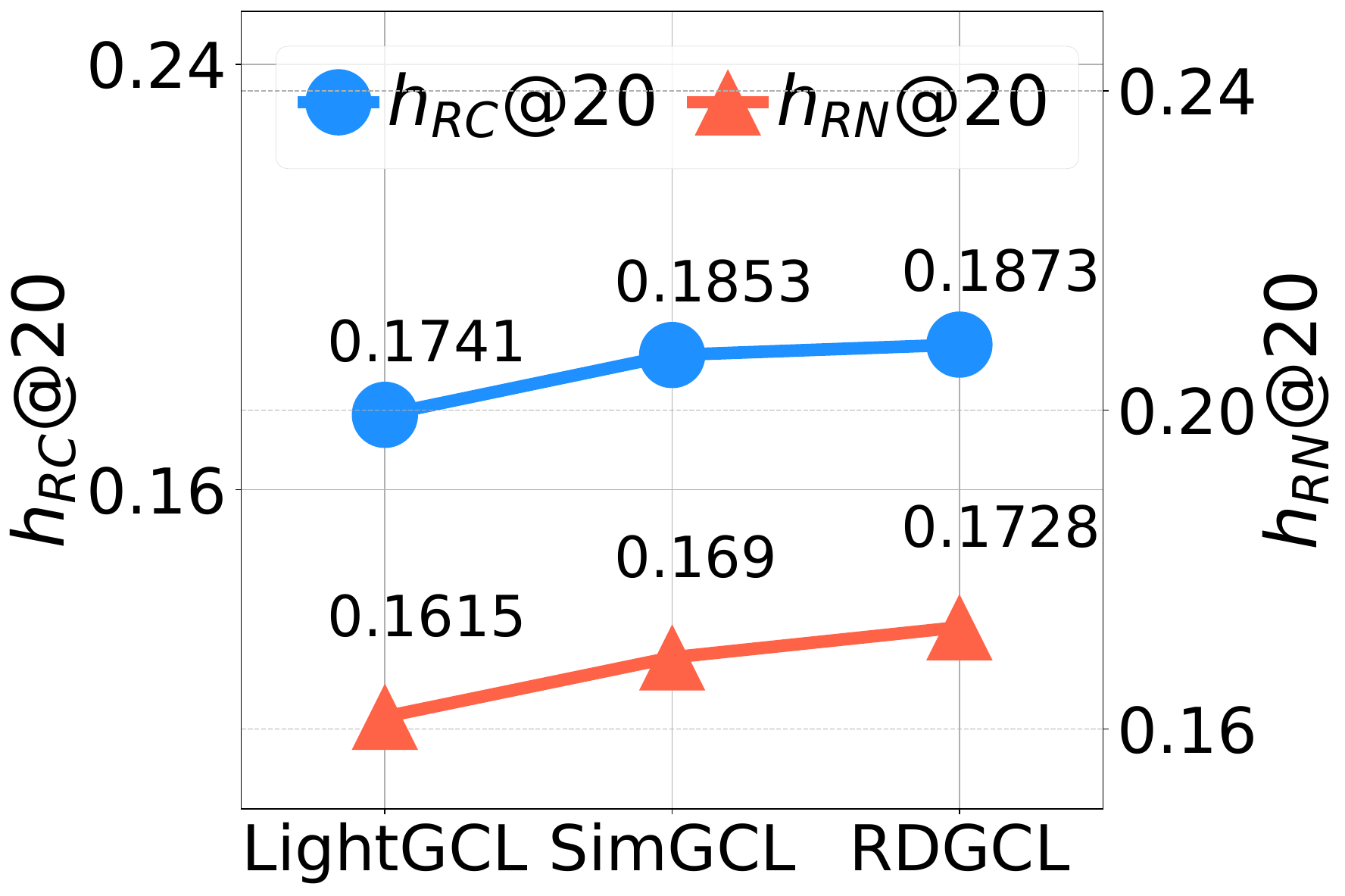}}
    \subfigure[Gowalla]{\includegraphics[width=0.49\columnwidth]{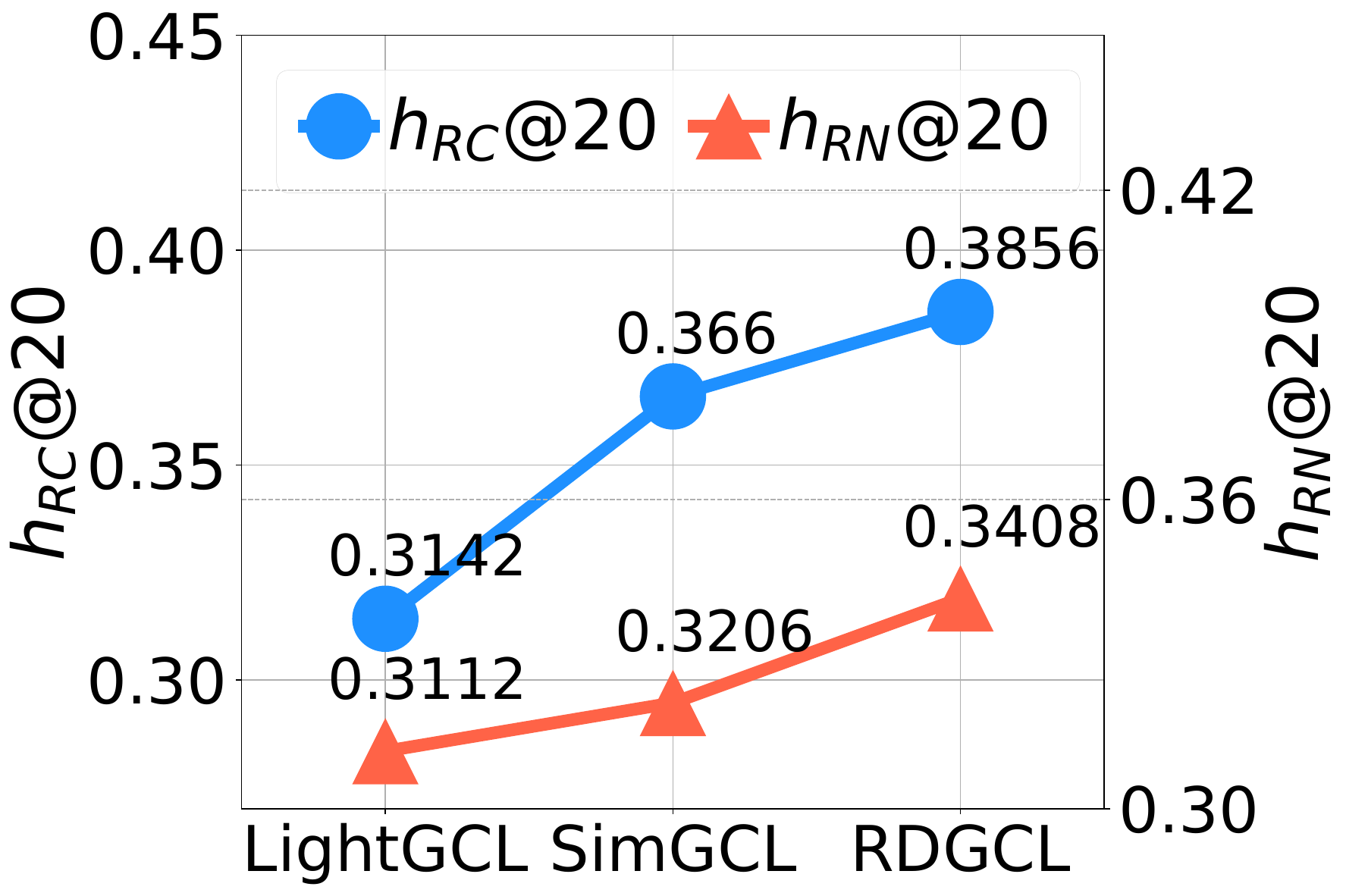}}
    \caption{Trade-off among the three metrics}
    \label{fig:diversity}
\end{figure}

As shown in Fig.~\ref{fig:diversity}, LightGCL has low $h_{\text{RC}}@20$ and $h_{\text{RN}}@20$, while RDGCL has the highest balanced performance. In the case of SimGCL, both metrics are lower than ours. This result suggests that RDGCL is a balanced design with improved accuracy and diversity.

\subsection{Robustness to Sparsity and Popularity Bias}\label{sec:robustness}
\begin{figure}[t!]
    \centering
    \subfigure[Yelp]{\includegraphics[width=0.49\columnwidth]{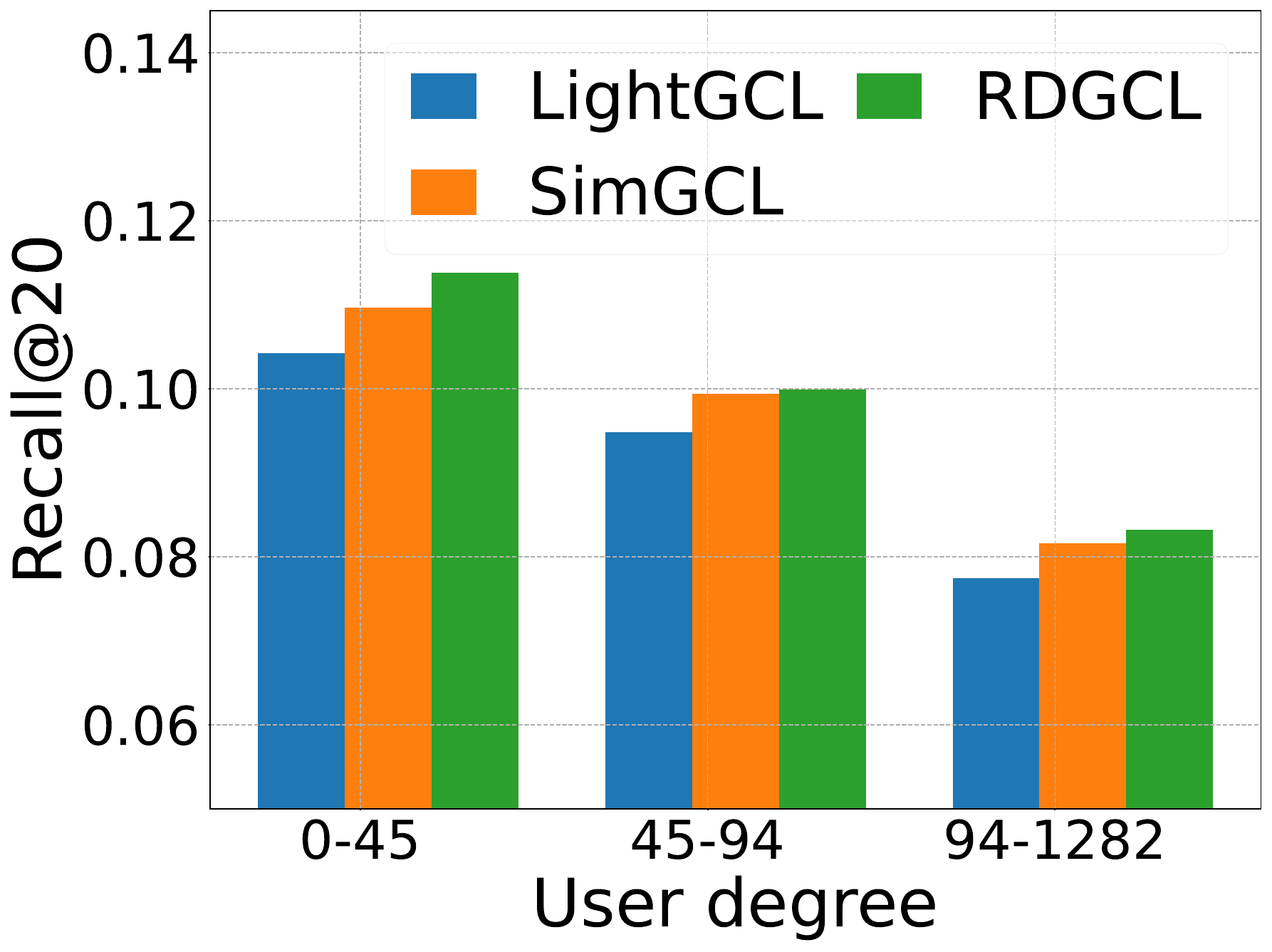}}
    \subfigure[Gowalla]{\includegraphics[width=0.49\columnwidth]{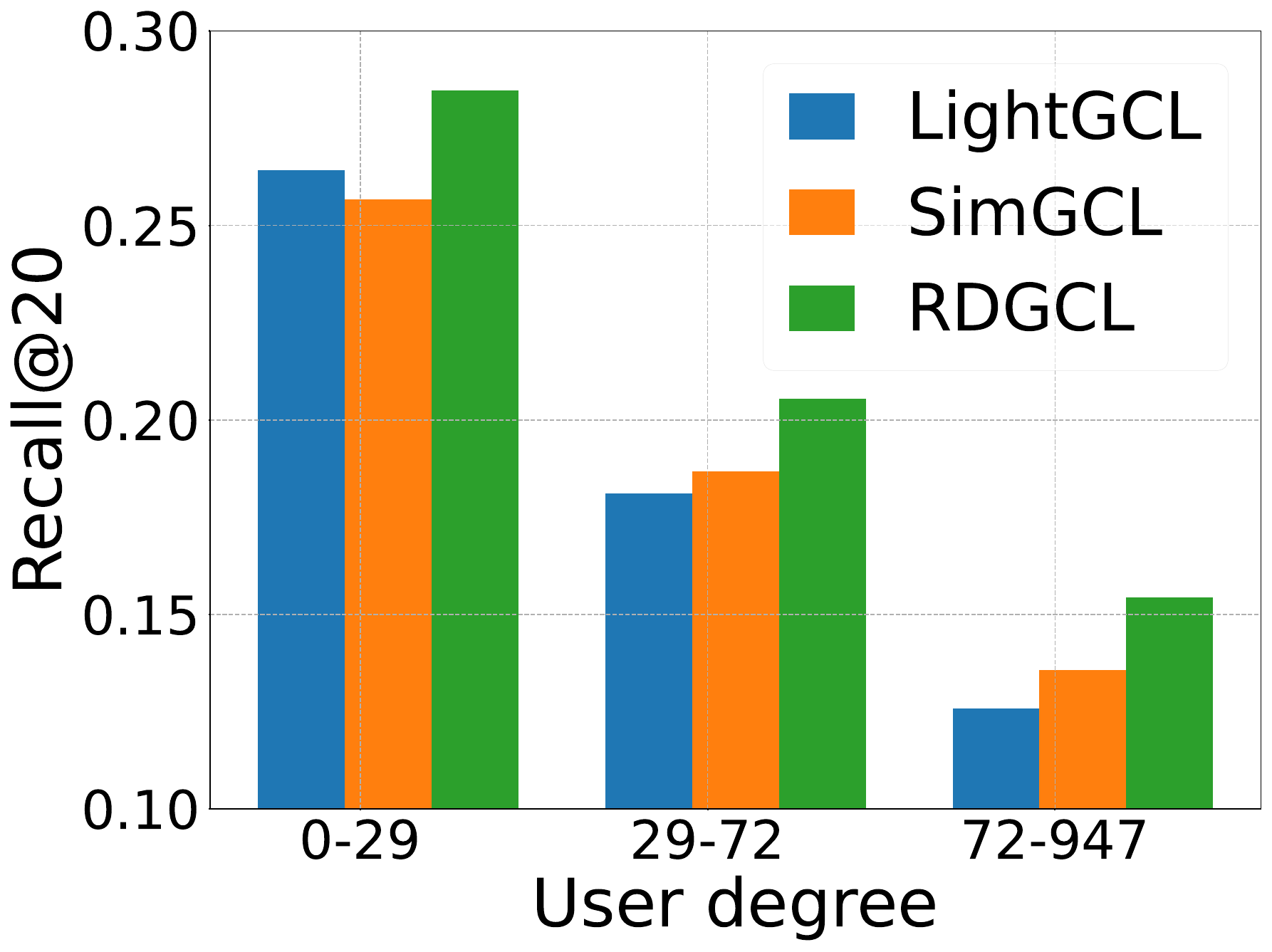}}
    \caption{Performance on users of different sparsity degrees}
    \label{fig:user_sparsity}
\end{figure}

To measure the robustness for sparse user groups, we divide users into three groups and measure Recall@20 for each group. The users were classified into three groups by interaction degree: the bottom 80\%, from the bottom 80\% to 95\%, and the top 5\%. As shown in Fig.~\ref{fig:user_sparsity}, RDGCL consistently outperforms the other baselines for all groups of users. Specifically, RDGCL shows good accuracy for extremely sparse user groups ($<29$) in Gowalla.

Furthermore, we study the robustness to the item popularity bias with our method and the baselines. We divide items into three groups based on their degree of interaction and measure Recall@20 on each group $g$. Following SGL~\cite{Wu2021SGL}, we use the decomposed Recall@$k$, defined as $\text{Recall@}k^{(g)}=\frac{|(l^u_{rec})^{(g)} \cap l^u_{test}|}{|l^u_{test}|}$.
$l^u_{rec}$ represents the candidate items in the top-$k$ recommendation list and $l^u_{test}$ represents the relevant items in the test set for user $u$. 

\begin{figure}[t!]
    \centering
    \subfigure[Yelp]{\includegraphics[width=0.49\columnwidth]{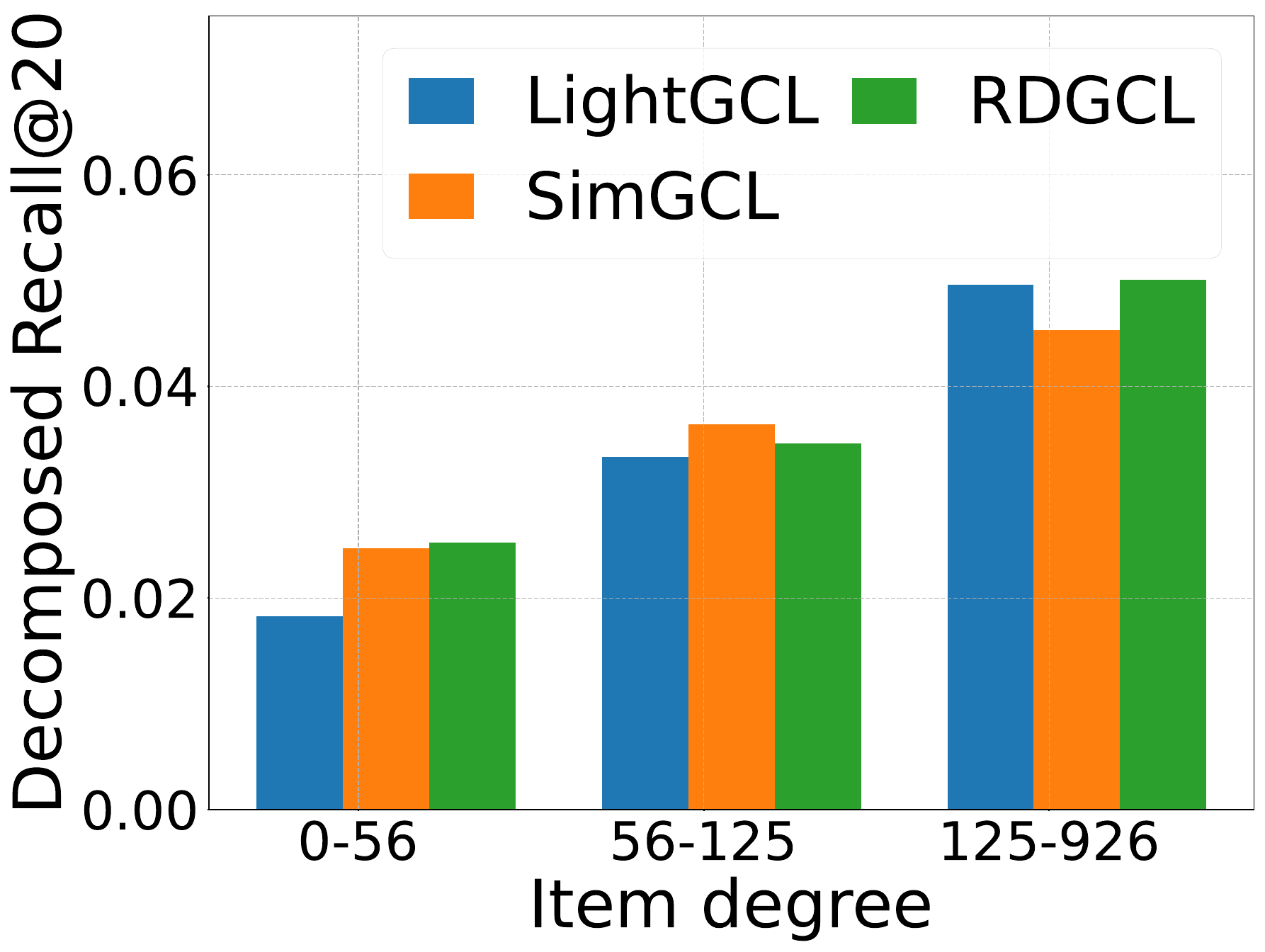}}
    \subfigure[Gowalla]{\includegraphics[width=0.49\columnwidth]{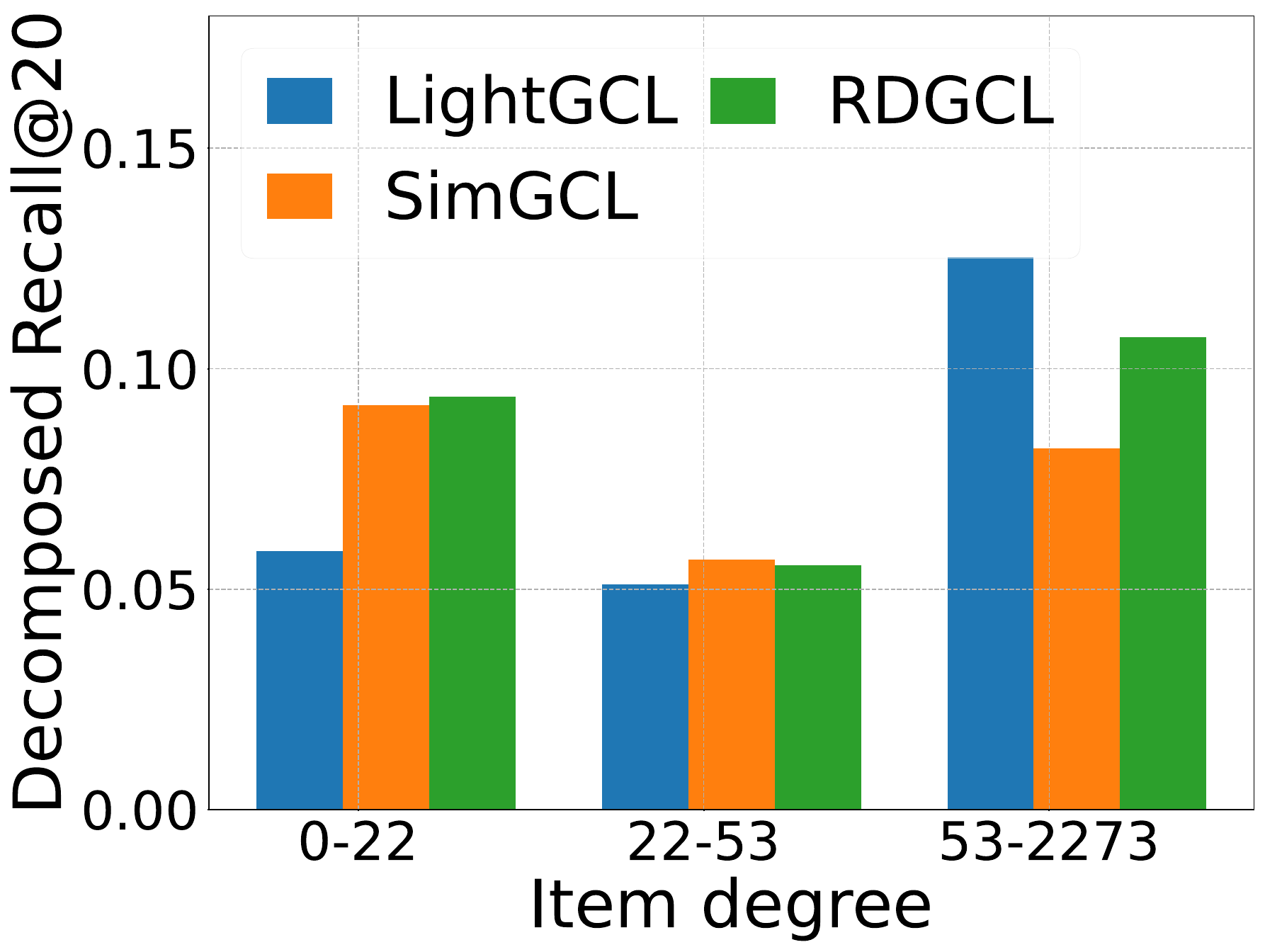}}
    \caption{RDGCL's ability to mitigate the popularity bias}
    \label{fig:popularity_bias}
\end{figure}

Fig.~\ref{fig:popularity_bias} shows that the accuracy of our model is higher for the item group with a low interaction degree compared to other baselines. This indicates that our model recommends long-tail items and has the ability to alleviate popularity bias.

\subsection{Robustness to Noise Interactions}\label{sec:noise}
\begin{figure}[h!]
    \centering
    \subfigure[Yelp]{\includegraphics[width=0.49\linewidth]{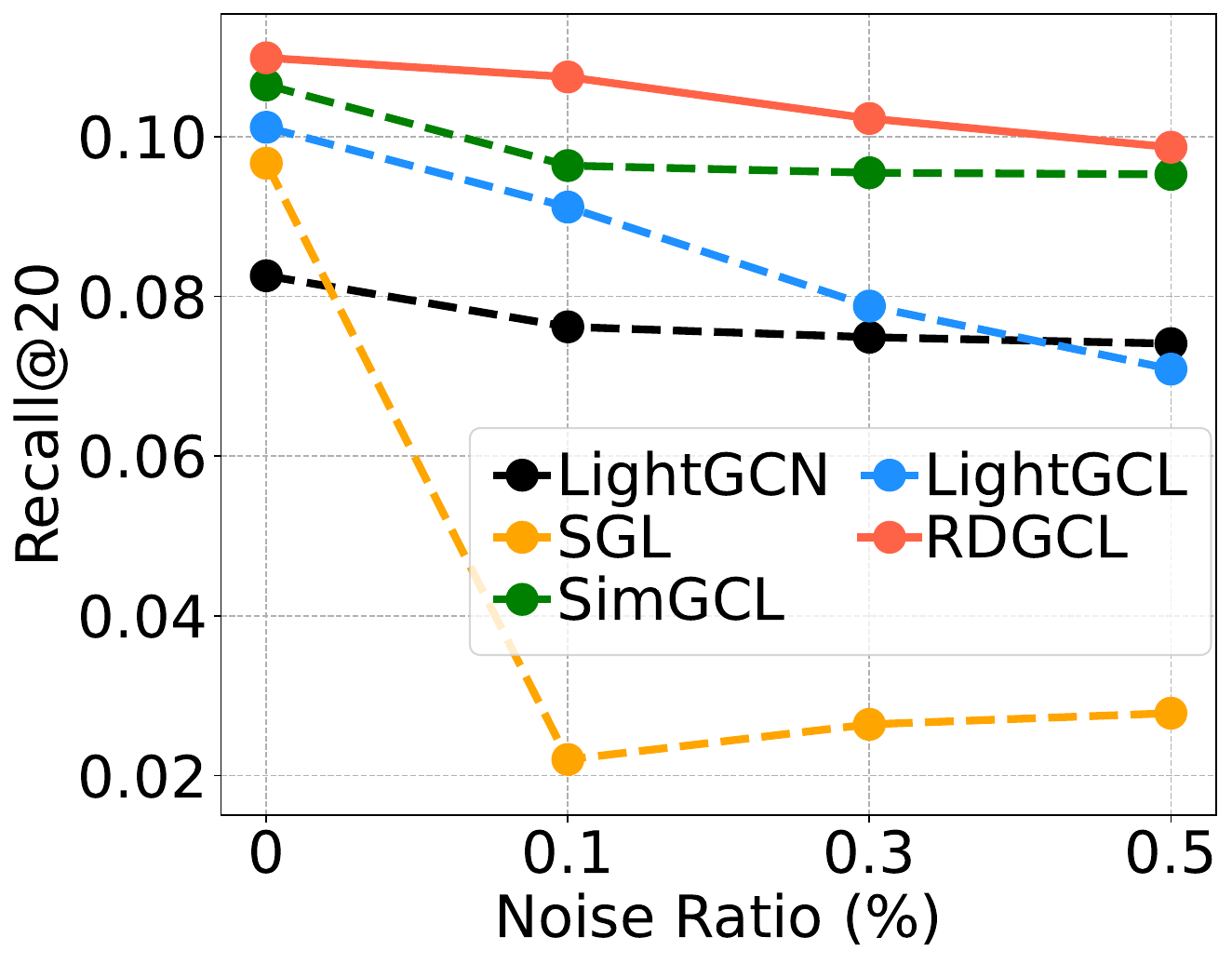}}
    \subfigure[Gowalla]{\includegraphics[width=0.49\linewidth]{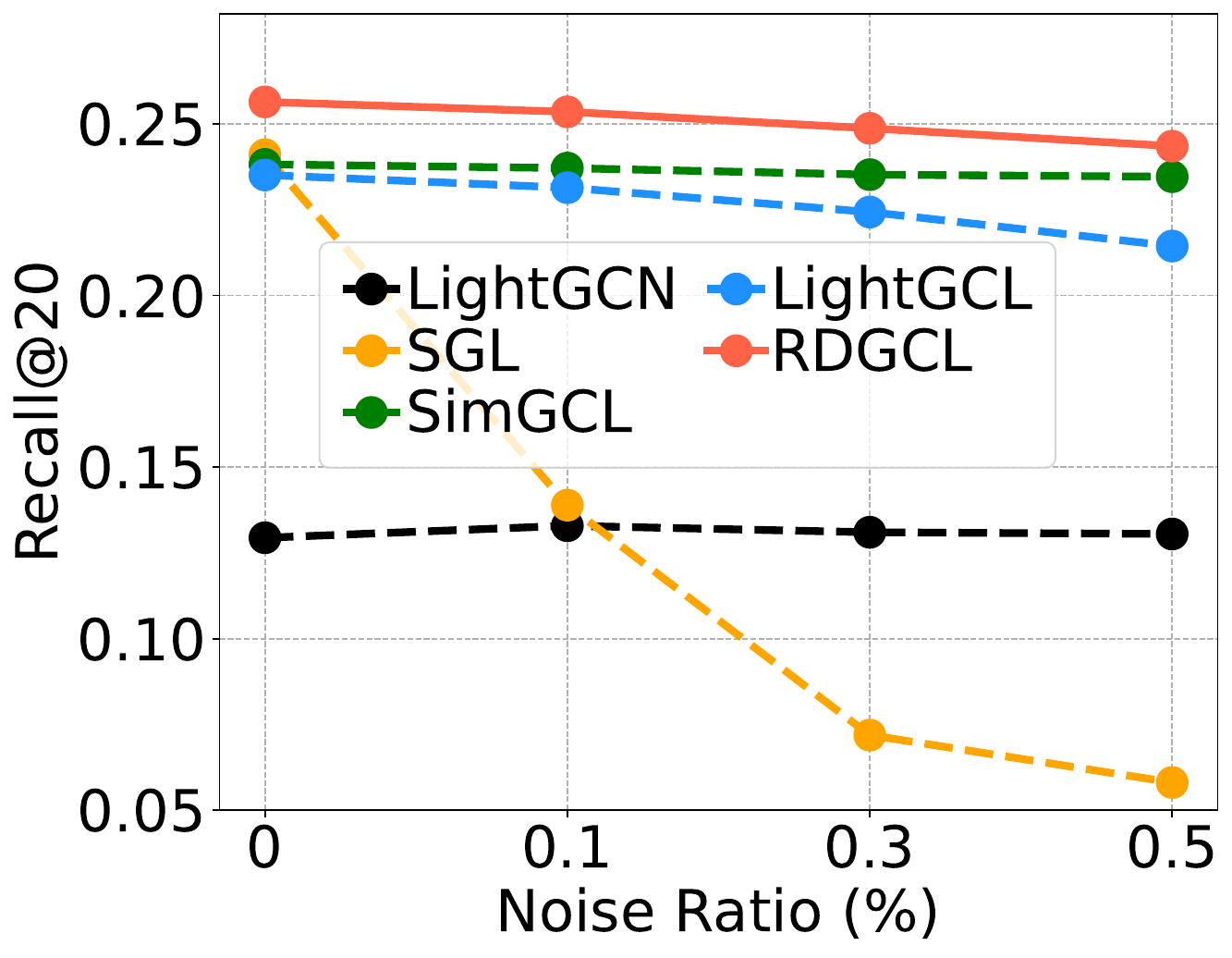}}
    \caption{Recall@20 w.r.t. noise ratio}
    \label{fig:noise}
\end{figure}

We evaluate the robustness of RDGCL and baselines to noise in user-item interactions. We add noise interactions randomly to the train dataset, train on the noisy train dataset, and evaluate it on the test dataset. We conduct by setting the percentage of random noise added to 0.1\%, 0.3\%, and 0.5\% of the total number of interactions. Fig.~\ref{fig:noise} shows the performance of the models with respect to the noise ratio.

In the case of Yelp, the performance of SGL and LightGCL tends to decrease due to added noise interactions. In particular, SGL appears to be less robust to noise. However, we can see that our RDGCL and SimGCL still show high recommendation performance compared to other models.

\subsection{Ablation Studies}
As ablation study models, we define 5 models:
i) ``RDGCL-EB'' contrasts $\mathbf{E}(T)$ and $\mathbf{B}^{CL}$, ii) ``RDGCL-ES'' contrasts $\mathbf{E}(T)$ and $\mathbf{S}^{CL}$, iii) ``w/o CL'' removes $\mathcal{L}_{CL}$,  iv) ``Only Diffusion'' uses only the diffusion term in Eq.~\eqref{eq:rd}, and v) ``Only Reaction'' uses only the reaction term in Eq.~\eqref{eq:rd}.

Table~\ref{tab:ablation} shows the results on Yelp and Gowalla.
RDGCL-ES outperforms RDGCL-EB, indicating the importance of the reaction process. 
The model without $\mathcal{L}_{CL}$ shows a performance drop, underlining the role of CL in enhancing accuracy. Interestingly, ``Only Diffusion" performs between SimGCL and LightGCL on Yelp, but still performs worse than RDGCL. ``Only Reaction'' fails to achieve complete alignment with RDGCL, highlighting the necessity of balancing both diffusion and reaction processes. These results validate the importance of each component in RDGCL.

\begin{table}[h]
    \small
    \centering
    \setlength{\tabcolsep}{1.5pt}
    \begin{tabular}{l cc cc}\toprule
        \multirow{2}{*}{Model} & \multicolumn{2}{c}{Yelp} & \multicolumn{2}{c}{Gowalla}\\\cmidrule(lr){2-3}\cmidrule(lr){4-5}
                             & Recall@20 & NDCG@20 & Recall@20 & NDCG@20\\\midrule
        RDGCL    & 0.1099 & 0.0939  & 0.2564 & 0.1549\\\cmidrule(lr){1-5}
        RDGCL-EB & 0.1024 & 0.0871  & 0.2495 & 0.1524\\
        RDGCL-ES & 0.1094 & 0.0935  & 0.2531 & 0.1530\\
        w/o CL   & 0.0950 & 0.0794  & 0.1565 & 0.0929\\
        Only Diffusion & 0.1034 & 0.0878  & 0.2359 & 0.1388\\
        Only Reaction  & 0.0875 & 0.0728  & 0.2266 & 0.1328\\
        \bottomrule
    \end{tabular}
    \caption{Ablation studies on RDGCL}
    \label{tab:ablation}
\end{table}

\subsection{Empirical Evaluations for Oversmoothing}\label{sec:over}
\begin{figure}[t]
    \centering
    \subfigure[Yelp]{\includegraphics[width=0.48\columnwidth]{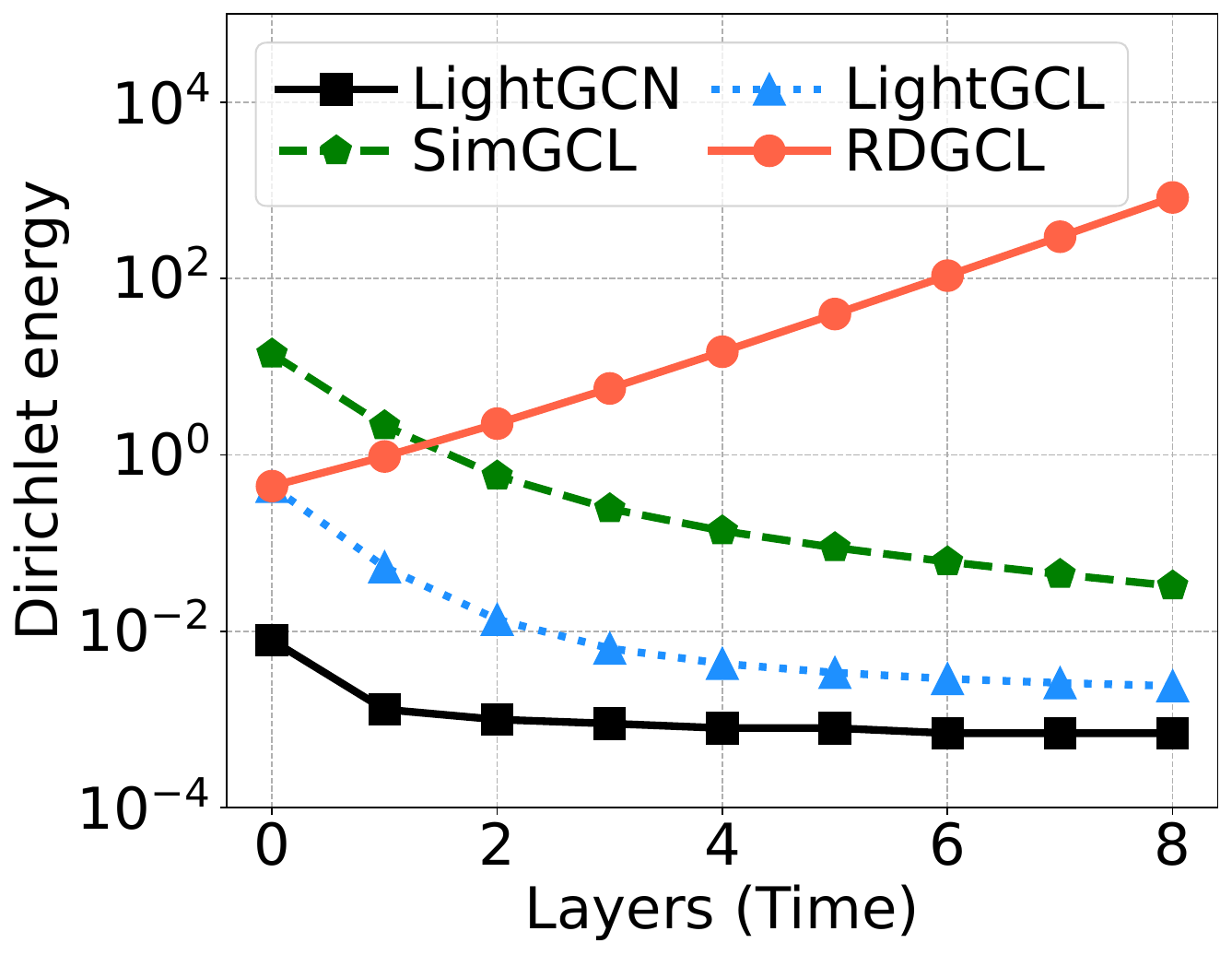}}
    \subfigure[Gowalla]{\includegraphics[width=0.48\columnwidth]{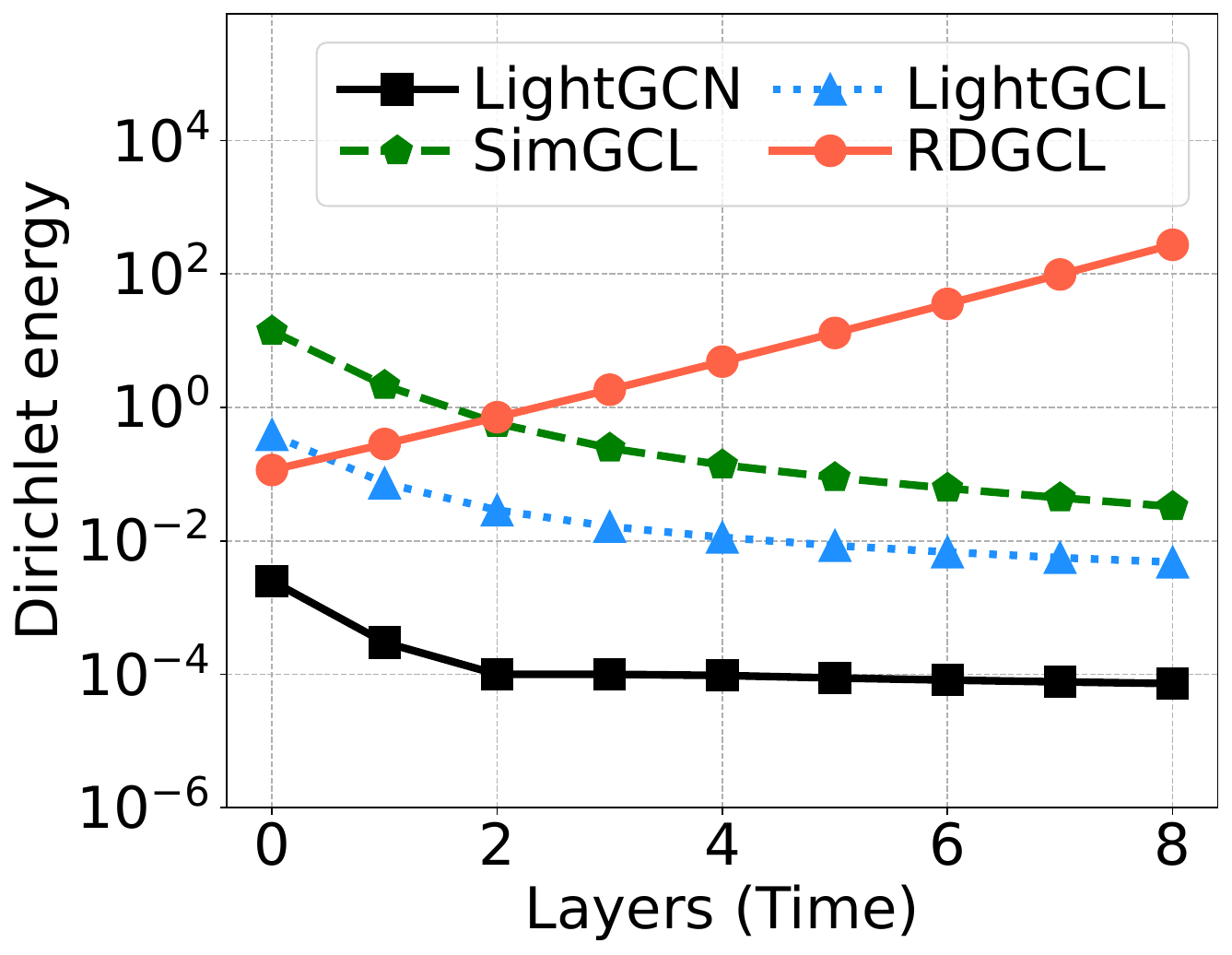}}
    \caption{Evolution of the Dirichlet energy}
    \label{fig:energy}
\end{figure}

We analyze the oversmoothing~\cite{oono2020oversmoothing} in RDGCL and existing graph-based methods. 
Oversmoothing occurs when user/item node feature similarities converge to a constant value as the number of GCN layers increases. 
To measure oversmoothing, we use Dirichlet energy~\cite{rusch2023survey}. Fig.~\ref{fig:energy} shows the Dirichlet energy of final embeddings for various methods as the number of layers, while other methods show decrease of energy. This provides the resilience of RDGCL to oversmoothing, which can be attributed to its distinctive combination of low-pass and high-pass filters.

\subsection{Sensitivity Analyses}
We conduct sensitivity analyses for RDGCL on key hyperparameters: the terminal integral time $T$, the temperature $\tau$, the regularization weight for InfoNCE loss $\lambda_1$, and $\alpha$.
The results for $T$ and $\alpha$ are presented in the main text, while the analyses for $\tau$ and $\lambda_1$ are in \emph{Appendix~\ref{app:sens}}.

\subsubsection{Sensitivity to $T$} 
\begin{figure}[t]
    \centering
    \subfigure[Yelp]{\includegraphics[width=0.46\columnwidth]{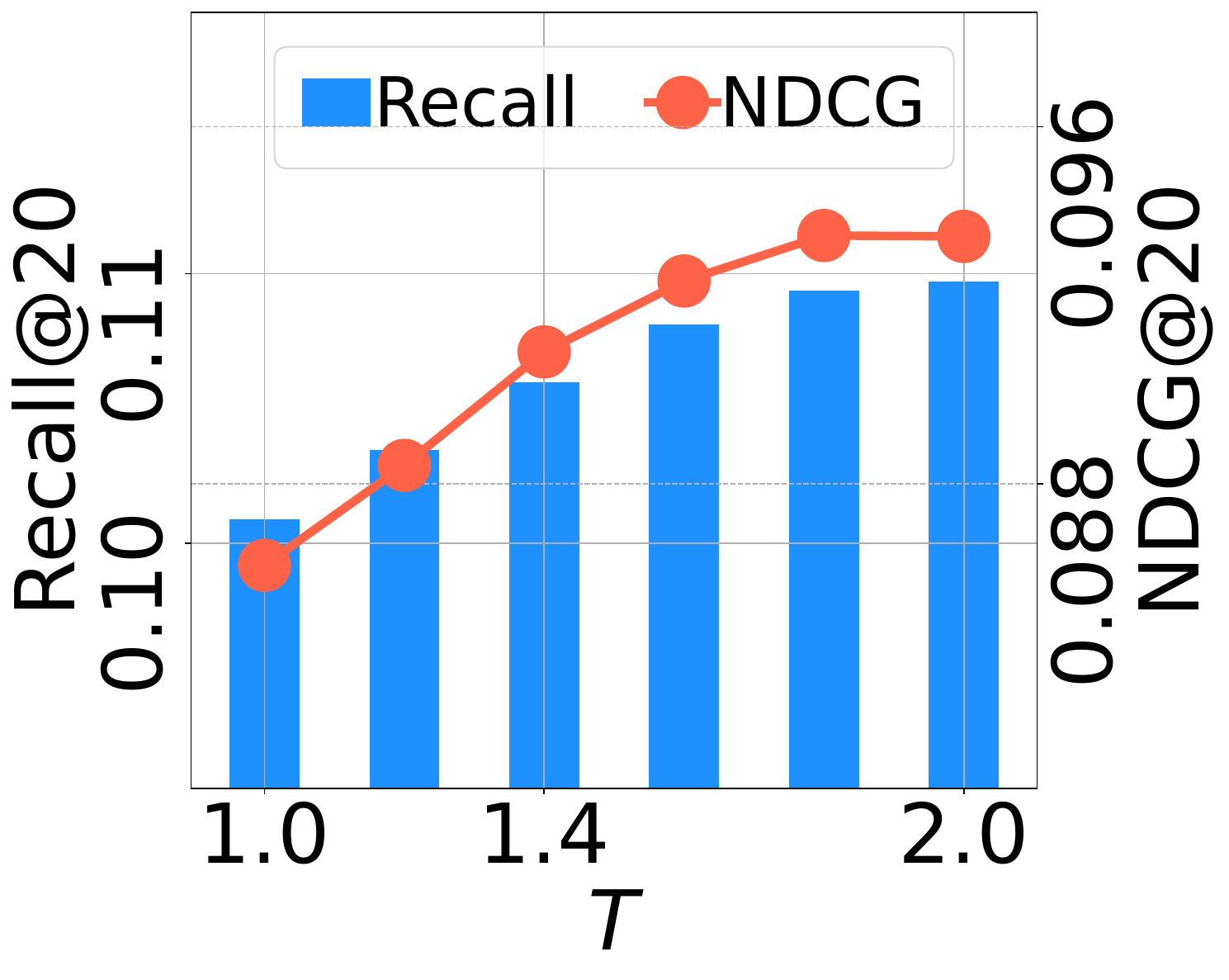}}
    \subfigure[Gowalla]{\includegraphics[width=0.46\columnwidth]{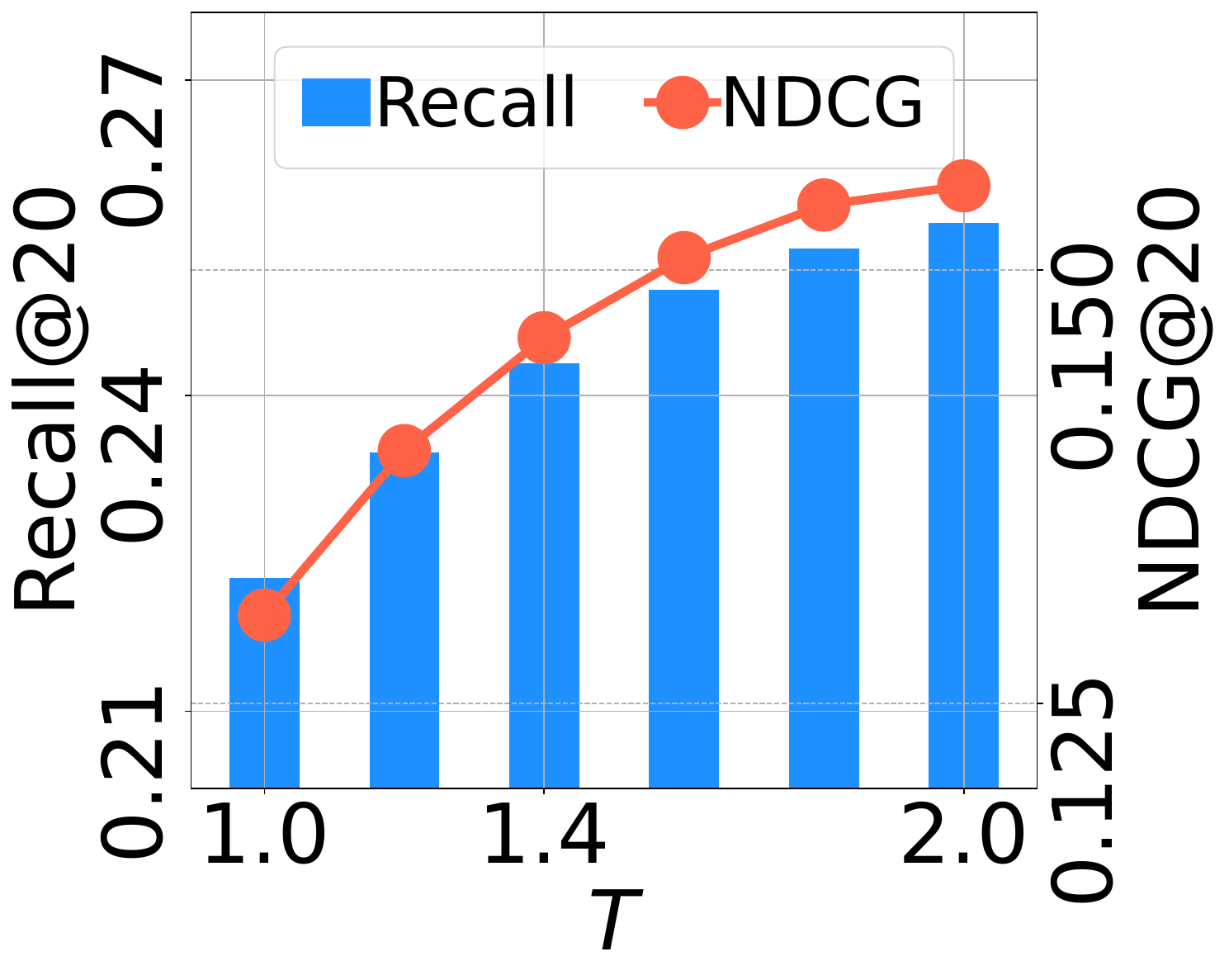}}
    \caption{Sensitivity on $T$}
    \label{fig:sens_T}
\end{figure}

We test the impact of the terminal integral time $T$ of the reaction-diffusion process on our model, and the results are shown in Fig.~\ref{fig:sens_T}. $T$ close to 2 yields best results for both Yelp and Gowalla.

\subsubsection{Sensitivity to $\alpha$}
\begin{figure}[t]
    \centering
    \subfigure[Yelp]{\includegraphics[width=0.46\columnwidth]{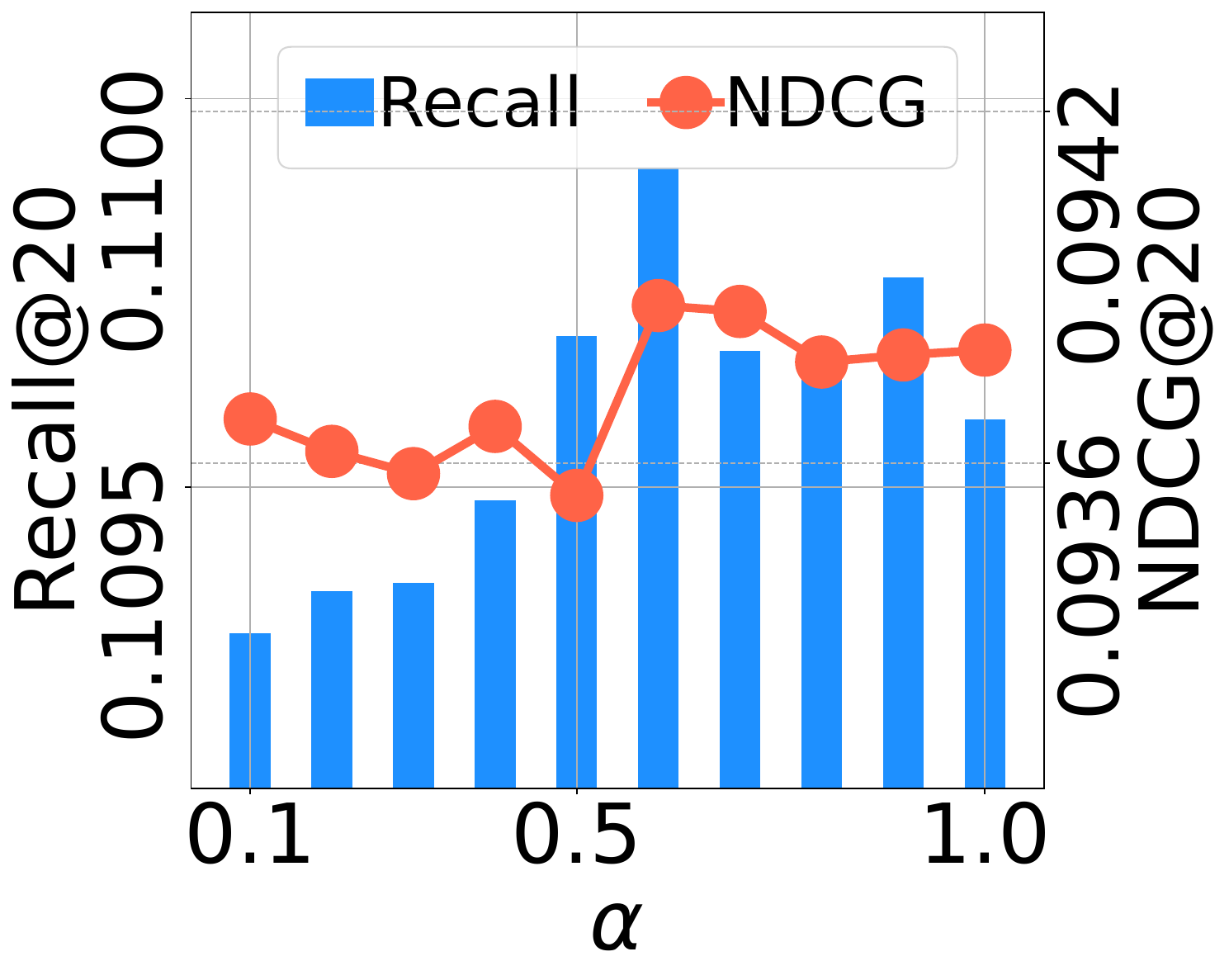}}
    \subfigure[Gowalla]{\includegraphics[width=0.46\columnwidth]{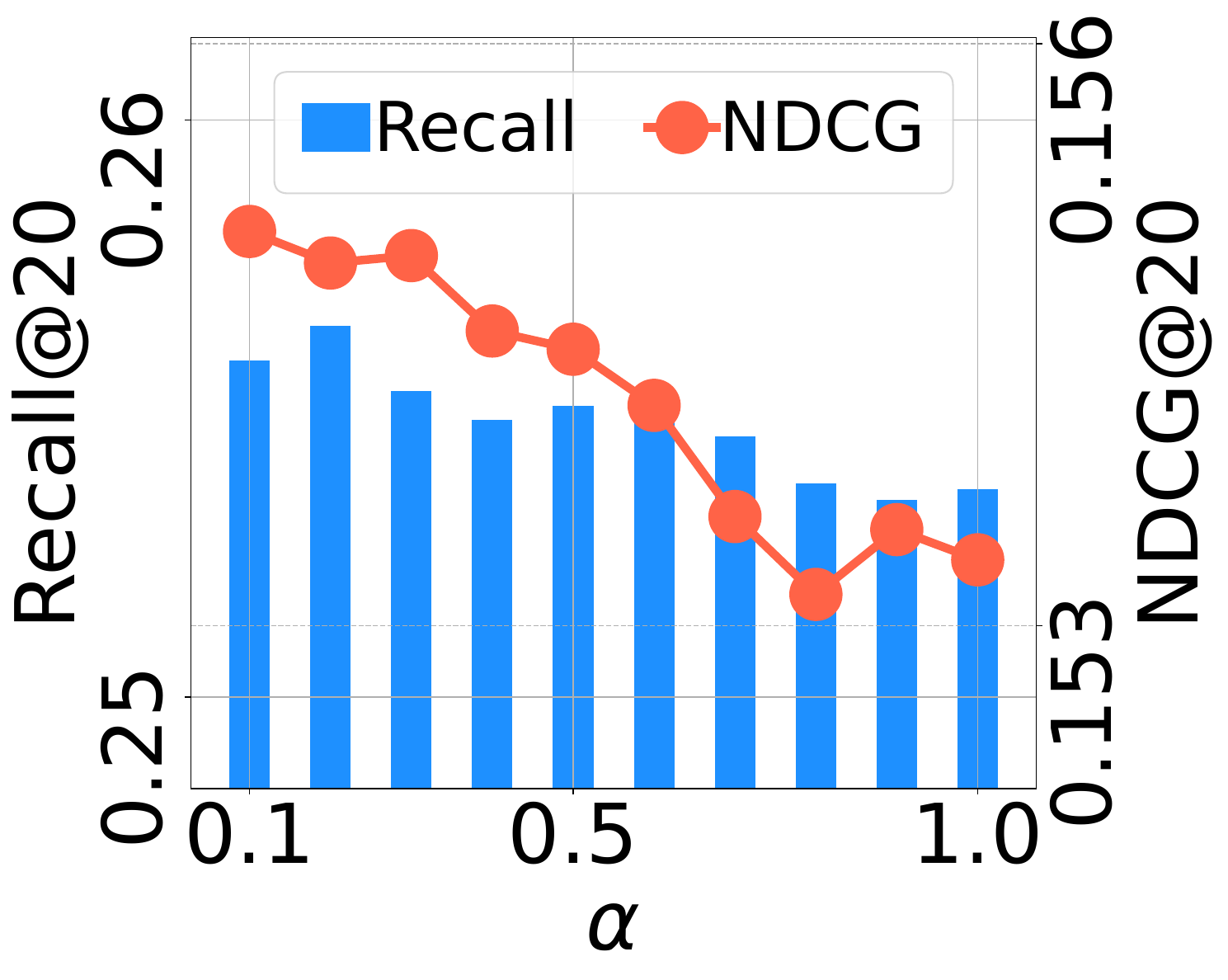}}
    \caption{Sensitivity on $\alpha$}
    \label{fig:sens_alpha}
\end{figure}

Since the intensity of the high-pass filter is determined by $\alpha$, we test how our RDGCL accuracy changes by varying the reaction rate coefficient, $\alpha$, in Fig.~\ref{fig:sens_alpha}. For Yelp, the best performance is shown when $\alpha$ is near 0.6, while for Gowalla, the optimal coefficient is 0.2. It shows that the effect of CL is maximized at an optimal reaction rate coefficient for each dataset.

\subsection{Empirical Runtime Complexity}
We also report our actual running time during training in Table~\ref{tab:runtime}. Our method does not involve augmentations on graph structures, and its single pipeline for CL is not separated from the main channel, unlike SGL and SimGCL. 
However, since RDGCL performs the reaction process, it has faster or comparable running times for some cases compared to XSimGCL and LightGCL. Nevertheless, our proposed method is still effective since it outperforms both XSimGCL and LightGCL by large margins in accuracy.

\begin{table}[t!]
    \centering
    \setlength{\tabcolsep}{1.5pt}
    \small
    \begin{tabular}{l ccccc}\toprule
        \multirow{2}{*}{Model} & \multirow{2}{*}{Yelp} & \multirow{2}{*}{Gowalla} & Amazon- & Amazon- & \multirow{2}{*}{Tmall} \\
                     & & & Electronics & CDs          &  \\\midrule
        SGL          & 44.91$s$ &  44.48$s$ & 19.58$s$& 0.26$s$ & 102.55$s$\\
        SimGCL       & 29.13$s$ &  50.02$s$ & 17.46$s$& 0.25$s$ & 154.08$s$\\
        XSimGCL      & 20.40$s$ &  38.37$s$ & 14.45$s$& 0.21$s$ &  89.05$s$\\
        LightGCL     & 19.94$s$ &  34.61$s$ & 11.81$s$& 0.20$s$ &  90.59$s$\\
        RDGCL        & 20.32$s$ &  34.89$s$ & 12.11$s$& 0.20$s$ & 93.31$s$\\ \bottomrule
    \end{tabular}
    \caption{Efficiency comparison, w.r.t. the average training time per epoch}
    \label{tab:runtime}
\end{table}

\section{Conclusion \& Limitation}
We presented a novel approach called RDGCL for CF. It uses both the diffusion equation for low-pass filtering and the reaction equation for high-pass filtering in its design and CL training method. To the best of our knowledge, RDGCL is the first to adopt the reaction-diffusion equation for CL-based CF. 
In addition, our method is more efficient than other CL-based methods due to its \emph{single pass} architecture.
We demonstrated that RDGCL outperforms 14 baselines on 5 datasets and achieves the best-balanced performance among the recall, coverage and novelty. 

Our limitation is that our reaction process requires another matrix multiplication computation in one layer.
However, our single-pass architecture allows it to be efficient or comparable in runtime terms to other CL-based CF methods. Extending the reaction process to a more efficient one leaves for future work.
\clearpage
\bibliography{ref}

\begin{thebibliography}{73}
\providecommand{\natexlab}[1]{#1}

\bibitem[{Allen and Cahn(1979)}]{allen1979microscopic}
Allen, S.~M.; and Cahn, J.~W. 1979.
\newblock A microscopic theory for antiphase boundary motion and its application to antiphase domain coarsening.
\newblock \emph{Acta metallurgica}, 27(6): 1085--1095.

\bibitem[{Bayer, Kaufhold, and Reuter(2022)}]{bayer2022survey}
Bayer, M.; Kaufhold, M.-A.; and Reuter, C. 2022.
\newblock A survey on data augmentation for text classification.
\newblock \emph{ACM Computing Surveys}, 55(7): 1--39.

\bibitem[{Cai et~al.(2023)Cai, Huang, Xia, and Ren}]{cai2023lightgcl}
Cai, X.; Huang, C.; Xia, L.; and Ren, X. 2023.
\newblock LightGCL: Simple Yet Effective Graph Contrastive Learning for Recommendation.
\newblock In \emph{ICLR}.

\bibitem[{Chamberlain et~al.(2021)Chamberlain, Rowbottom, Goronova, Webb, Rossi, and Bronstein}]{chamberlain2021grand}
Chamberlain, B.~P.; Rowbottom, J.; Goronova, M.; Webb, S.; Rossi, E.; and Bronstein, M.~M. 2021.
\newblock {GRAND}: Graph Neural Diffusion.
\newblock In \emph{ICML}.

\bibitem[{Chen et~al.(2020{\natexlab{a}})Chen, Wu, Hong, Zhang, and Wang}]{chen20LRGCCF}
Chen, L.; Wu, L.; Hong, R.; Zhang, K.; and Wang, M. 2020{\natexlab{a}}.
\newblock Revisiting Graph Based Collaborative Filtering: A Linear Residual Graph Convolutional Network Approach.
\newblock In \emph{AAAI}.

\bibitem[{Chen et~al.(2018)Chen, Rubanova, Bettencourt, and Duvenaud}]{chen2018NODE}
Chen, R. T.~Q.; Rubanova, Y.; Bettencourt, J.; and Duvenaud, D.~K. 2018.
\newblock Neural Ordinary Differential Equations.
\newblock In \emph{NeurIPS}.

\bibitem[{Chen et~al.(2020{\natexlab{b}})Chen, Kornblith, Norouzi, and Hinton}]{chen2020simple}
Chen, T.; Kornblith, S.; Norouzi, M.; and Hinton, G. 2020{\natexlab{b}}.
\newblock A Simple Framework for Contrastive Learning of Visual Representations.
\newblock \emph{arXiv preprint arXiv:2002.05709}.

\bibitem[{Chen et~al.(2020{\natexlab{c}})Chen, Kornblith, Swersky, Norouzi, and Hinton}]{chen2020big}
Chen, T.; Kornblith, S.; Swersky, K.; Norouzi, M.; and Hinton, G. 2020{\natexlab{c}}.
\newblock Big Self-Supervised Models are Strong Semi-Supervised Learners.
\newblock \emph{arXiv preprint arXiv:2006.10029}.

\bibitem[{Chen, Luo, and Li(2021)}]{chen2021intriguing}
Chen, T.; Luo, C.; and Li, L. 2021.
\newblock Intriguing properties of contrastive losses.
\newblock \emph{NeruIPS}, 34: 11834--11845.

\bibitem[{Chen and Pock(2016)}]{chen2016trainable}
Chen, Y.; and Pock, T. 2016.
\newblock Trainable nonlinear reaction diffusion: A flexible framework for fast and effective image restoration.
\newblock \emph{IEEE transactions on pattern analysis and machine intelligence}, 39(6): 1256--1272.

\bibitem[{Choi et~al.(2023{\natexlab{a}})Choi, Hong, Park, and Cho}]{choi2023bspm}
Choi, J.; Hong, S.; Park, N.; and Cho, S.-B. 2023{\natexlab{a}}.
\newblock Blurring-Sharpening Process Models for Collaborative Filtering.
\newblock In \emph{SIGIR}.

\bibitem[{Choi et~al.(2023{\natexlab{b}})Choi, Hong, Park, and Cho}]{choi2023gread}
Choi, J.; Hong, S.; Park, N.; and Cho, S.-B. 2023{\natexlab{b}}.
\newblock GREAD: Graph Neural Reaction-Diffusion Networks.
\newblock In \emph{ICML}.

\bibitem[{Choi, Jeon, and Park(2021)}]{choi2021ltocf}
Choi, J.; Jeon, J.; and Park, N. 2021.
\newblock LT-OCF: Learnable-Time ODE-based Collaborative Filtering.
\newblock In \emph{CIKM}.

\bibitem[{Chuang et~al.(2022)Chuang, Dangovski, Luo, Zhang, Chang, Soljacic, Li, Yih, Kim, and Glass}]{chuang2022diffcse}
Chuang, Y.-S.; Dangovski, R.; Luo, H.; Zhang, Y.; Chang, S.; Soljacic, M.; Li, S.-W.; Yih, W.-t.; Kim, Y.; and Glass, J. 2022.
\newblock {DiffCSE}: Difference-based Contrastive Learning for Sentence Embeddings.
\newblock In \emph{NAACL}.

\bibitem[{Chung(1997)}]{chung1997spectral}
Chung, F.~R. 1997.
\newblock \emph{Spectral graph theory}, volume~92.
\newblock American Mathematical Soc.

\bibitem[{Du et~al.(2023)Du, Yuan, Zhao, Qu, Zhuang, Liu, Liu, and Sheng}]{du2023fearec}
Du, X.; Yuan, H.; Zhao, P.; Qu, J.; Zhuang, F.; Liu, G.; Liu, Y.; and Sheng, V.~S. 2023.
\newblock Frequency Enhanced Hybrid Attention Network for Sequential Recommendation.
\newblock In \emph{SIGIR}, 78--88.

\bibitem[{Ekambaram(2013)}]{Ekambaram2013lens}
Ekambaram, V.~N. 2013.
\newblock Graph Structured Data Viewed Through a Fourier Lens.
\newblock \emph{University of California, Berkeley}.

\bibitem[{Fan et~al.(2022)Fan, Liu, Jin, Zhao, Tang, and Li}]{fan2022GTN}
Fan, W.; Liu, X.; Jin, W.; Zhao, X.; Tang, J.; and Li, Q. 2022.
\newblock Graph Trend Filtering Networks for Recommendation.
\newblock In \emph{SIGIR}, 112–121.

\bibitem[{Fisher(1937)}]{fisher1937wave}
Fisher, R.~A. 1937.
\newblock The wave of advance of advantageous genes.
\newblock \emph{Annals of eugenics}, 7(4): 355--369.

\bibitem[{Harper and Konstan(2015)}]{harper2015movielens}
Harper, F.~M.; and Konstan, J.~A. 2015.
\newblock The movielens datasets: History and context.
\newblock \emph{Acm transactions on interactive intelligent systems (tiis)}, 5(4): 1--19.

\bibitem[{Hassani and Khasahmadi(2020)}]{hassani2020contrastive}
Hassani, K.; and Khasahmadi, A.~H. 2020.
\newblock Contrastive multi-view representation learning on graphs.
\newblock In \emph{ICML}, 4116--4126. PMLR.

\bibitem[{He et~al.(2020{\natexlab{a}})He, Fan, Wu, Xie, and Girshick}]{he2020momentum}
He, K.; Fan, H.; Wu, Y.; Xie, S.; and Girshick, R. 2020{\natexlab{a}}.
\newblock Momentum contrast for unsupervised visual representation learning.
\newblock In \emph{CVPR}, 9729--9738.

\bibitem[{He et~al.(2020{\natexlab{b}})He, Deng, Wang, Li, Zhang, and Wang}]{He20LightGCN}
He, X.; Deng, K.; Wang, X.; Li, Y.; Zhang, Y.; and Wang, M. 2020{\natexlab{b}}.
\newblock LightGCN: Simplifying and Powering Graph Convolution Network for Recommendation.
\newblock In \emph{SIGIR}.

\bibitem[{Herlocker et~al.(2004)Herlocker, Konstan, Terveen, and Riedl}]{Herlocker2004Coverage}
Herlocker, J.~L.; Konstan, J.~A.; Terveen, L.~G.; and Riedl, J.~T. 2004.
\newblock Evaluating Collaborative Filtering Recommender Systems.
\newblock \emph{ACM Trans. Inf. Syst.}, 22(1): 5–53.

\bibitem[{Hong et~al.(2022)Hong, Jo, Kook, Jung, Wi, Park, and Cho}]{hong2022timekit}
Hong, S.; Jo, M.; Kook, S.; Jung, J.; Wi, H.; Park, N.; and Cho, S.-B. 2022.
\newblock {TimeKit}: A Time-series Forecasting-based Upgrade Kit for Collaborative Filtering.
\newblock In \emph{2022 IEEE International Conference on Big Data (Big Data)}, 565--574. IEEE.

\bibitem[{Hu et~al.(2022)Hu, Qian, Fang, and Xu}]{hu2022mgdcf}
Hu, J.; Qian, S.; Fang, Q.; and Xu, C. 2022.
\newblock MGDCF: Distance Learning via Markov Graph Diffusion for Neural Collaborative Filtering.
\newblock \emph{arXiv preprint arXiv: Arxiv-2204.02338}.

\bibitem[{Jaiswal et~al.(2020)Jaiswal, Babu, Zadeh, Banerjee, and Makedon}]{jaiswal2020survey}
Jaiswal, A.; Babu, A.~R.; Zadeh, M.~Z.; Banerjee, D.; and Makedon, F. 2020.
\newblock A survey on contrastive self-supervised learning.
\newblock \emph{Technologies}, 9(1): 2.

\bibitem[{Jing et~al.(2023)Jing, Zhu, Zang, and Wang}]{jing2023survey}
Jing, M.; Zhu, Y.; Zang, T.; and Wang, K. 2023.
\newblock Contrastive Self-supervised Learning in Recommender Systems: A Survey.
\newblock \emph{arXiv preprint arXiv: Arxiv-2303.09902}.

\bibitem[{Khosla et~al.(2020)Khosla, Teterwak, Wang, Sarna, Tian, Isola, Maschinot, Liu, and Krishnan}]{khosla2020supervised}
Khosla, P.; Teterwak, P.; Wang, C.; Sarna, A.; Tian, Y.; Isola, P.; Maschinot, A.; Liu, C.; and Krishnan, D. 2020.
\newblock Supervised contrastive learning.
\newblock \emph{NeurIPS}, 33: 18661--18673.

\bibitem[{Kondo and Miura(2010)}]{kondo2010reaction}
Kondo, S.; and Miura, T. 2010.
\newblock Reaction-diffusion model as a framework for understanding biological pattern formation.
\newblock \emph{science}, 329(5999): 1616--1620.

\bibitem[{Kong et~al.(2022)Kong, Kim, Jeon, Choi, Lee, Park, and Kim}]{kong2022hmlet}
Kong, T.; Kim, T.; Jeon, J.; Choi, J.; Lee, Y.-C.; Park, N.; and Kim, S.-W. 2022.
\newblock Linear, or Non-Linear, That is the Question!
\newblock In \emph{WSDM}, 517--525.

\bibitem[{Lee, Kim, and Lee(2018)}]{lee2018goccf}
Lee, Y.-C.; Kim, S.-W.; and Lee, D. 2018.
\newblock {gOCCF}: Graph-theoretic one-class collaborative filtering based on uninteresting items.
\newblock In \emph{AAAI}, volume~32.

\bibitem[{Li et~al.(2023)Li, Guo, Zhu, Li, Wang, and Chen}]{li2023sgccl}
Li, B.; Guo, T.; Zhu, X.; Li, Q.; Wang, Y.; and Chen, F. 2023.
\newblock SGCCL: siamese graph contrastive consensus learning for personalized recommendation.
\newblock In \emph{WSDM}, 589--597.

\bibitem[{Lin et~al.(2022)Lin, Tian, Hou, and Zhao}]{lin2022ncl}
Lin, Z.; Tian, C.; Hou, Y.; and Zhao, W.~X. 2022.
\newblock Improving graph collaborative filtering with neighborhood-enriched contrastive learning.
\newblock In \emph{TheWebConf (former WWW)}, 2320--2329.

\bibitem[{Liu et~al.(2021{\natexlab{a}})Liu, Cheng, Zhu, Gao, and Nie}]{liu2021IMP-GCN}
Liu, F.; Cheng, Z.; Zhu, L.; Gao, Z.; and Nie, L. 2021{\natexlab{a}}.
\newblock Interest-Aware Message-Passing GCN for Recommendation.
\newblock In \emph{TheWebConf (former WWW)}, 1296–1305.

\bibitem[{Liu et~al.(2021{\natexlab{b}})Liu, Zhang, Hou, Mian, Wang, Zhang, and Tang}]{liu2021self}
Liu, X.; Zhang, F.; Hou, Z.; Mian, L.; Wang, Z.; Zhang, J.; and Tang, J. 2021{\natexlab{b}}.
\newblock Self-supervised learning: Generative or contrastive.
\newblock \emph{IEEE Transactions on Knowledge and Data Engineering}, 35(1): 857--876.

\bibitem[{Mao et~al.(2021{\natexlab{a}})Mao, Zhu, Wang, Dai, Dong, Xiao, and He}]{mao2021simplex}
Mao, K.; Zhu, J.; Wang, J.; Dai, Q.; Dong, Z.; Xiao, X.; and He, X. 2021{\natexlab{a}}.
\newblock SimpleX: A Simple and Strong Baseline for Collaborative Filtering.
\newblock In \emph{CIKM}, 1243–1252.

\bibitem[{Mao et~al.(2021{\natexlab{b}})Mao, Zhu, Xiao, Lu, Wang, and He}]{Mao21UltraGCN}
Mao, K.; Zhu, J.; Xiao, X.; Lu, B.; Wang, Z.; and He, X. 2021{\natexlab{b}}.
\newblock UltraGCN: Ultra Simplification of Graph Convolutional Networks for Recommendation.
\newblock In \emph{CIKM}.

\bibitem[{Nt and Maehara(2019)}]{nt2019revisiting}
Nt, H.; and Maehara, T. 2019.
\newblock Revisiting graph neural networks: All we have is low-pass filters.
\newblock \emph{arXiv preprint arXiv:1905.09550}.

\bibitem[{Oono and Suzuki(2020)}]{oono2020oversmoothing}
Oono, K.; and Suzuki, T. 2020.
\newblock Graph neural networks exponentially lose expressive power for node classification.
\newblock In \emph{ICLR}.

\bibitem[{Oord, Li, and Vinyals(2018)}]{oord2018infonce}
Oord, A. v.~d.; Li, Y.; and Vinyals, O. 2018.
\newblock Representation learning with contrastive predictive coding.
\newblock \emph{arXiv preprint arXiv:1807.03748}.

\bibitem[{Peng, Sugiyama, and Mine(2022)}]{Peng2022Less}
Peng, S.; Sugiyama, K.; and Mine, T. 2022.
\newblock Less is More: Reweighting Important Spectral Graph Features for Recommendation.
\newblock In \emph{SIGIR}, 1273–1282.

\bibitem[{Pinar(2021)}]{pinar2021analytical}
Pinar, Z. 2021.
\newblock An Analytical Studies of the Reaction-Diffusion Systems of Chemical Reactions.
\newblock \emph{International Journal of Applied and Computational Mathematics}, 7(3): 81.

\bibitem[{Plonka and Ma(2008)}]{plonka2008nonlinear}
Plonka, G.; and Ma, J. 2008.
\newblock Nonlinear regularized reaction-diffusion filters for denoising of images with textures.
\newblock \emph{IEEE Transactions on Image Processing}, 17(8): 1283--1294.

\bibitem[{Poli et~al.(2019)Poli, Massaroli, Park, Yamashita, Asama, and Park}]{poli2019gde}
Poli, M.; Massaroli, S.; Park, J.; Yamashita, A.; Asama, H.; and Park, J. 2019.
\newblock Graph neural ordinary differential equations.
\newblock \emph{arXiv preprint arXiv:1911.07532}.

\bibitem[{Qiu et~al.(2020)Qiu, Chen, Dong, Zhang, Yang, Ding, Wang, and Tang}]{qiu2020gcc}
Qiu, J.; Chen, Q.; Dong, Y.; Zhang, J.; Yang, H.; Ding, M.; Wang, K.; and Tang, J. 2020.
\newblock {GCC}: Graph contrastive coding for graph neural network pre-training.
\newblock In \emph{KDD}, 1150--1160.

\bibitem[{Radford et~al.(2021)Radford, Kim, Hallacy, Ramesh, Goh, Agarwal, Sastry, Askell, Mishkin, Clark et~al.}]{radford2021learning}
Radford, A.; Kim, J.~W.; Hallacy, C.; Ramesh, A.; Goh, G.; Agarwal, S.; Sastry, G.; Askell, A.; Mishkin, P.; Clark, J.; et~al. 2021.
\newblock Learning transferable visual models from natural language supervision.
\newblock In \emph{ICML}, 8748--8763. PMLR.

\bibitem[{Rusch, Bronstein, and Mishra(2023)}]{rusch2023survey}
Rusch, T.~K.; Bronstein, M.~M.; and Mishra, S. 2023.
\newblock A Survey on Oversmoothing in Graph Neural Networks.
\newblock \emph{arXiv preprint arXiv: Arxiv-2303.10993}.

\bibitem[{Shen et~al.(2021)Shen, Wu, Zhang, Shan, Zhang, Letaief, and Li}]{Shen21GFCF}
Shen, Y.; Wu, Y.; Zhang, Y.; Shan, C.; Zhang, J.; Letaief, B.~K.; and Li, D. 2021.
\newblock How Powerful is Graph Convolution for Recommendation?
\newblock In \emph{CIKM}.

\bibitem[{Shin et~al.(2023)Shin, Choi, Wi, and Park}]{shin2023attentive}
Shin, Y.; Choi, J.; Wi, H.; and Park, N. 2023.
\newblock An Attentive Inductive Bias for Sequential Recommendation Beyond the Self-Attention.
\newblock \emph{arXiv preprint arXiv:2312.10325}.

\bibitem[{Singer, Shkolnisky, and Nadler(2009)}]{singer2009diffusion}
Singer, A.; Shkolnisky, Y.; and Nadler, B. 2009.
\newblock Diffusion interpretation of nonlocal neighborhood filters for signal denoising.
\newblock \emph{SIAM Journal on Imaging Sciences}, 2(1): 118--139.

\bibitem[{Sullivan and Feinn(2012)}]{sullivan2012using}
Sullivan, G.~M.; and Feinn, R. 2012.
\newblock Using effect size—or why the P value is not enough.
\newblock \emph{Journal of graduate medical education}, 4(3): 279--282.

\bibitem[{Turing(1952)}]{alan1952morphogenesis}
Turing, A. 1952.
\newblock The chemical basis of morphogenesis.
\newblock \emph{Phil. Trans. R. Soc. Lond. B}.

\bibitem[{Turk(1991)}]{turk1991generating}
Turk, G. 1991.
\newblock Generating textures on arbitrary surfaces using reaction-diffusion.
\newblock \emph{Acm Siggraph Computer Graphics}, 25(4): 289--298.

\bibitem[{Wang et~al.(2019)Wang, He, Wang, Feng, and Chua}]{Wang19NGCF}
Wang, X.; He, X.; Wang, M.; Feng, F.; and Chua, T.-S. 2019.
\newblock Neural Graph Collaborative Filtering.
\newblock In \emph{SIGIR}.

\bibitem[{Wang et~al.(2021)Wang, Wang, Yang, and Lin}]{wang2021dgc}
Wang, Y.; Wang, Y.; Yang, J.; and Lin, Z. 2021.
\newblock Dissecting the Diffusion Process in Linear Graph Convolutional Networks.
\newblock In \emph{NeurIPS}.

\bibitem[{Wang et~al.(2023)Wang, Yi, Liu, Wang, and Jin}]{wang2023acmp}
Wang, Y.; Yi, K.; Liu, X.; Wang, Y.~G.; and Jin, S. 2023.
\newblock {ACMP}: Allen-Cahn Message Passing for Graph Neural Networks with Particle Phase Transition.
\newblock In \emph{ICLR}.

\bibitem[{Witkin and Kass(1991)}]{witkin1991reaction}
Witkin, A.; and Kass, M. 1991.
\newblock Reaction-diffusion textures.
\newblock In \emph{Proceedings of the 18th annual conference on computer graphics and interactive techniques}, 299--308.

\bibitem[{Wu et~al.(2021)Wu, Wang, Feng, He, Chen, Lian, and Xie}]{Wu2021SGL}
Wu, J.; Wang, X.; Feng, F.; He, X.; Chen, L.; Lian, J.; and Xie, X. 2021.
\newblock Self-Supervised Graph Learning for Recommendation.
\newblock In \emph{SIGIR}, 726–735.

\bibitem[{Xhonneux, Qu, and Tang(2020)}]{xhonneux2019CGNN}
Xhonneux, L.-P. A.~C.; Qu, M.; and Tang, J. 2020.
\newblock Continuous Graph Neural Networks.
\newblock In \emph{ICML}.

\bibitem[{Xia et~al.(2022{\natexlab{a}})Xia, Wu, Chen, Hu, and Li}]{xia2022SimGRACE}
Xia, J.; Wu, L.; Chen, J.; Hu, B.; and Li, S.~Z. 2022{\natexlab{a}}.
\newblock SimGRACE: A Simple Framework for Graph Contrastive Learning without Data Augmentation.
\newblock In \emph{TheWebConf (former WWW)}.

\bibitem[{Xia et~al.(2022{\natexlab{b}})Xia, Huang, Xu, Zhao, Yin, and Huang}]{xia2022HCCF}
Xia, L.; Huang, C.; Xu, Y.; Zhao, J.; Yin, D.; and Huang, J. 2022{\natexlab{b}}.
\newblock Hypergraph contrastive collaborative filtering.
\newblock In \emph{SIGIR}, 70--79.

\bibitem[{Xia, Huang, and Zhang(2022)}]{xia2022SHT}
Xia, L.; Huang, C.; and Zhang, C. 2022.
\newblock Self-supervised hypergraph transformer for recommender systems.
\newblock In \emph{KDD}, 2100--2109.

\bibitem[{Xu et~al.(2023)Xu, Wang, Wang, Guo, Fan, Tian, and Wang}]{xu2023simdcl}
Xu, Y.; Wang, Z.; Wang, Z.; Guo, Y.; Fan, R.; Tian, H.; and Wang, X. 2023.
\newblock SimDCL: dropout-based simple graph contrastive learning for recommendation.
\newblock \emph{Complex \& Intelligent Systems}, 1--13.

\bibitem[{You et~al.(2020)You, Chen, Sui, Chen, Wang, and Shen}]{you2020GraphCL}
You, Y.; Chen, T.; Sui, Y.; Chen, T.; Wang, Z.; and Shen, Y. 2020.
\newblock Graph contrastive learning with augmentations.
\newblock \emph{NeurIPS}, 33: 5812--5823.

\bibitem[{Yu et~al.(2018)Yu, Gao, Li, Yin, and Liu}]{yu2018ifbpr}
Yu, J.; Gao, M.; Li, J.; Yin, H.; and Liu, H. 2018.
\newblock Adaptive implicit friends identification over heterogeneous network for social recommendation.
\newblock In \emph{CIKM}, 357--366.

\bibitem[{Yu et~al.(2022{\natexlab{a}})Yu, Xia, Chen, zhen Cui, Hung, and Yin}]{yu2022xsimgcl}
Yu, J.; Xia, X.; Chen, T.; zhen Cui, L.; Hung, N. Q.~V.; and Yin, H. 2022{\natexlab{a}}.
\newblock XSimGCL: Towards Extremely Simple Graph Contrastive Learning for Recommendation.
\newblock \emph{arXiv preprint arXiv:2209.02544}.

\bibitem[{Yu et~al.(2022{\natexlab{b}})Yu, Yin, Xia, Chen, Cui, and Nguyen}]{yu2022SimGCL}
Yu, J.; Yin, H.; Xia, X.; Chen, T.; Cui, L.; and Nguyen, Q. V.~H. 2022{\natexlab{b}}.
\newblock Are graph augmentations necessary? simple graph contrastive learning for recommendation.
\newblock In \emph{SIGIR}, 1294--1303.

\bibitem[{Zhao et~al.(2022)Zhao, Wu, Liang, Chen, Zhang, Deng, Wang, Shen, Lv, and Wu}]{zhao2022investigating}
Zhao, M.; Wu, L.; Liang, Y.; Chen, L.; Zhang, J.; Deng, Q.; Wang, K.; Shen, X.; Lv, T.; and Wu, R. 2022.
\newblock Investigating accuracy-novelty performance for graph-based collaborative filtering.
\newblock In \emph{SIGIR}, 50--59.

\bibitem[{Zhou et~al.(2023)Zhou, Chen, Dong, Zha, Zhou, and Huang}]{zhou2023adaptive}
Zhou, H.; Chen, H.; Dong, J.; Zha, D.; Zhou, C.; and Huang, X. 2023.
\newblock Adaptive popularity debiasing aggregator for graph collaborative filtering.
\newblock In \emph{SIGIR}, 7--17.

\bibitem[{Zhou et~al.(2010)Zhou, Kuscsik, Liu, Medo, Wakeling, and Zhang}]{zhou2010solving}
Zhou, T.; Kuscsik, Z.; Liu, J.-G.; Medo, M.; Wakeling, J.~R.; and Zhang, Y.-C. 2010.
\newblock Solving the apparent diversity-accuracy dilemma of recommender systems.
\newblock \emph{Proceedings of the National Academy of Sciences}, 107(10): 4511--4515.

\bibitem[{Zhu et~al.(2020)Zhu, Xu, Yu, Liu, Wu, and Wang}]{zhu2020deep}
Zhu, Y.; Xu, Y.; Yu, F.; Liu, Q.; Wu, S.; and Wang, L. 2020.
\newblock Deep graph contrastive representation learning.
\newblock \emph{arXiv preprint arXiv:2006.04131}.

\bibitem[{Zhu et~al.(2021)Zhu, Xu, Yu, Liu, Wu, and Wang}]{zhu2021GCA}
Zhu, Y.; Xu, Y.; Yu, F.; Liu, Q.; Wu, S.; and Wang, L. 2021.
\newblock Graph contrastive learning with adaptive augmentation.
\newblock In \emph{TheWebConf (former WWW)}, 2069--2080.

\end{thebibliography}
\clearpage

\appendix
{\huge \textbf{\textit{Appendix}}}

\section{Full Derivation of RDG Layer Filter}\label{app:rdg_derivation}
We provide the full derivation of the $2\tilde{\mathbf{A}}-\tilde{\mathbf{A}}^2$ filter applied in the RDG Layer under Euler discretization with $dt=1$, $T=1$, and $\alpha=1$:
\begin{align}
\begin{split}
    \mathbf{E}(T) &= \mathbf{E}(0) -\tilde{\mathbf{L}}\mathbf{E}(0) + \tilde{\mathbf{L}}\tilde{\mathbf{A}}\mathbf{E}(0)\\
         &= \mathbf{E}(0) + (\tilde{\mathbf{A}}-\mathbf{I})\mathbf{E}(0) + (\mathbf{I}-\tilde{\mathbf{A}}) \tilde{\mathbf{A}} \mathbf{E}(0)\\
         &= (2\tilde{\mathbf{A}} - \tilde{\mathbf{A}}^2) \mathbf{E}(0).
\end{split}
\end{align}
This derivation shows how the RDG layer applies the $2\tilde{\mathbf{A}}-\tilde{\mathbf{A}}^2$ filter to the initial embedding $\mathbf{E}(0)$, resulting in the final embedding $\mathbf{E}(T)$.

\section{Proof of Theorem~\ref{thrm:filter}}\label{app:proof}

\begin{proof}
Let $\gamma_i$ be an eigenvalue of $\tilde{\mathbf{A}}$. As noted in the paper, $\gamma_i \in (-1, 1]$ for normalized adjacency matrices. The frequency response of the $\tilde{\mathbf{A}}$ filter is $g_1(\gamma_i) = \gamma_i$, while the RDGCL filter has a frequency response of $g_2(\gamma_i) = 2\gamma_i - \gamma_i^2$.
To compare these filters, we analyze the difference of their responses:
\begin{align}
\begin{split}
\Delta g(\gamma_i) 
&= g_2(\gamma_i) - g_1(\gamma_i) \\
&= 2\gamma_i - \gamma_i^2 - \gamma_i \\
&= \gamma_i(1-\gamma_i).
\end{split}\nonumber
\end{align}
This difference function $\Delta f(\gamma_i)$ forms a parabola in the interval $[0,1]$, starting at zero when $\gamma_i = 0$, reaching a maximum value of $0.25$ at $\gamma_i = 0.5$, and returning to zero at $\gamma_i = 1$. This behavior is consistent with the findings of \citet{singer2009diffusion}.

In the context of graph signal processing with the normalized adjacency matrix $\tilde{\mathbf{A}}$, eigenvalues closer to 1 correspond to relatively higher-frequency signals, while those closer to 0 correspond to lower-frequency signals. The positive $\Delta g(\gamma_i)$ in $(0, 1)$ indicates that the RDGCL filter emphasizes higher frequency components more than the $\tilde{\mathbf{A}}$ filter in this range.

Formally, we can state:
\begin{align}
\forall \gamma_i \in (0, 1): \Delta g(\gamma_i) > 0. \nonumber
\end{align}

This property enables the RDGCL filter to selectively emphasize relatively higher-frequency signals while attenuating low-frequency signals. In contrast to the $\tilde{\mathbf{A}}$ filter, which acts as a low-pass filter, the RDGCL filter provides a more balanced frequency response.

Therefore, we conclude that the RDGCL filter enhances relatively higher frequency components compared to the $\tilde{\mathbf{A}}$ filter, particularly in the range $\gamma_i \in (0.5, 1)$, while still maintaining some low-pass characteristics. 
\end{proof}

This property enables RDGCL to maintain a balance between preserving higher frequency details and some degree of smoothing, potentially leading to better performance in tasks that require both local and global information processing.

\section{Details of Datasets}\label{app:data}
For Yelp and Gowalla, we use the data settings from~\citet{cai2023lightgcl}. For Amazon-Electronics and Amazon-CDs, we use the dataset settings used by~\citet{Mao21UltraGCN}.
We provide the detailed dataset statistics in Table~\ref{tab:data}.

\begin{table}[h]
    \small
    \centering
    \setlength{\tabcolsep}{3pt}
    \begin{tabular}{l cccc}\toprule
        Dataset     & \#Users & \#Items & \#Interactions & Density \\ \midrule
        Yelp        & 29,601  & 24,734  & 1,374,594       & 0.188\% \\ 
        Gowalla     & 50,821  & 57,440  & 1,302,695       & 0.045\% \\
        Amazon-Electronics & 1,435  & 1,522  & 35,931     & 1.645\% \\
        Amazon-CDs  & 43,169  & 35,648  & 777,426     & 0.051\% \\
        Tmall       & 47,939  & 41,390  & 2,619,389       & 0.132\% \\
        \bottomrule
    \end{tabular}
    \caption{Statistics of datasets}\label{tab:data}
\end{table}

\section{Details of Experimental Settings \& Hyperparameters}\label{app:exp_detail}
For our method, we test the following hyperparameters:
\begin{itemize}
    \item For solving the integral problem, we consider the Euler method;
    \item The number of steps $K$ is in $\{1, 2, 3\}$, and the terminal time $T$ is in $\{1.0,1.1,\cdots,3.0\}$; 
    \item The reaction rate coefficient $\alpha$ is in $\{0.1,\cdots,1.0\}$;
    \item For fair comparison, we set the embedding sizes for all methods to 256;
    \item The learning rate (lr) is in \{\num{1.0e-4}, \num{5.0e-4}, \num{1.0e-3}, \num{2.0e-3}, \num{3.0e-3}, \num{5.0e-3}, \num{1.0e-2}\};
    \item The temperature $\tau$ in the InfoNCE loss is in \{0.1, 0.2, 0.4, 0.6, 0.8, 1.0, 2.0, 5.0\};
    \item The regularization weight for the InfoNCE loss $\lambda_1$ is in $\{0.01, 0.1, 0.2, 0.3, 0.4, 0.5\}$;
    \item The regularization weight $\lambda_2$ is in $\{\num{1.0e-8},\num{1.0e-7},\num{1.0e-6},\num{1.0e-5},\num{5.0e-5}\}$.
\end{itemize}

We perform a grid search within the above search range to get the best set of configurations from each dataset, which are as follows:
\begin{itemize}
    \item In Yelp, $K=2$, $T=2$, $D=256$, $\text{lr}=\num{5e-04}$, $\alpha=0.6$, $\tau=0.1$, $\lambda_1=0.3$, and $\lambda_2=\num{5e-05}$. 
    \item In Gowalla, $K=2$, $T=2$, $D=256$, $\text{lr}=\num{1e-03}$, $\alpha=0.2$, $\tau=0.4$, $\lambda_1=0.5$, and $\lambda_2=1e-06$. 
    \item In Amazon-Electronics, $K=3$, $T=3$, $D=256$, $\text{lr}=\num{1e-03}$, $\alpha=0.1$, $\tau=5.0$, and $\lambda_1=0.01$.
    \item In Amazon-CDs, $K=2$, $T=2$, $D=256$, $\text{lr}=\num{1e-3}$, $\alpha=0.1$, $\tau=0.2$, $\lambda_1=0.2$, $\lambda_2=1e-7$.
    \item In Tmall, $K=2$, $T=2$,  $D=256$, $\text{lr}=\num{1e-3}$, $\alpha=0.6$, $\tau=0.7$, $\lambda_1=0.4$, $\lambda_2=1e-7$.
\end{itemize}

\pagebreak
\section{Statistical Test with Cohen's Effect Size}\label{app:cohen}
To evaluate the performance of RDGCL against the baseline models in all datasets, we apply Cohen's d to quantify the effect size~\cite{sullivan2012using}. This standardized measure of the difference between two means shows that RDGCL outperforms the baseline in almost all cases across all datasets. Specifically, the Cohen's d results indicate not only statistical significance, but also practical significance, indicating a real-world benefit of RDGCL over the baseline.

The effect size calculated using Cohen's d range is `small' for 0.2 and above, and `medium' for 0.5 and above. In Table~\ref{tab:cohen}, the results show that the improvement in RDGCL performance in terms of NDCG@20, a key metric for recommendation systems, is statistically significant. However, compared to SGL and LightGCL, it fails to reach 0.2 in terms of Recall@20 by a small margin. However, we believe that our RDGCL is an attractive choice for improving recommendation algorithms, not only because of the accuracy of its recommendations, but also because of the novelty and diversity of the items it recommends, as shown in Sec.~\ref{sec:trade}.

\begin{table}[t]
    \small
    \setlength{\tabcolsep}{2pt}
    \centering
    \begin{tabular}{l ccccc}\toprule
        \multirow{2}{*}{RDGCL v.s.} & \multicolumn{4}{c}{Cohen's d} \\\cmidrule(lr){2-5}
                   & Recall@20 & NDCG@20 & NDCG@20 & NDCG@40 \\\midrule
        LightGCN   &  0.81$\star$ & 1.20$\star$ & 0.78$\star$ & 1.17$\star$ \\
        LT-OCF     &  0.42$\star$ & 1.63$\star$ & 0.40$\ast$ & 0.62$\star$ \\
        HMLET      &  0.50$\star$ & 1.71$\star$ & 0.49$\star$ & 0.78$\star$ \\
        SGL        &  0.13\;\;    & 1.36$\star$ & 0.14\;\;    & 0.23$\ast$  \\
        SimGRACE   &  1.02$\star$ & 2.12$\star$ & 1.01$\star$ & 1.47$\star$ \\
        GCA        &  0.70$\star$ & 1.84$\star$ & 0.86$\star$ & 1.07$\star$ \\
        HCCF       &  0.88$\star$ & 1.88$\star$ & 0.85$\star$ & 1.04$\star$ \\
        SHT        &  0.80$\star$ & 1.93$\star$ & 0.77$\star$ & 1.13$\star$ \\
        SimGCL     &  0.15\;\;    & 1.37$\star$ & 0.15\;\;  & 0.27$\ast$  \\
        XSimGCL    &  0.39$\ast$  & 1.60$\star$ & 0.34$\ast$  & 0.56$\star$ \\
        LightGCL   &  0.16\;\;    & 1.37$\star$ & 0.15\;\;    & 0.22$\ast$  \\
        GFCF       &  0.36$\ast$  & 1.54$\star$ & 0.32$\ast$  & 0.48$\ast$  \\
        BSPM       &  0.28$\ast$  & 1.44$\star$ & 0.32$\ast$  & 0.39$\ast$  \\
        \bottomrule
    \end{tabular}
    \caption{The effect size using Cohen's d. $\ast$ means a small effect size greater than 0.2, and $\star$ means a medium effect size greater than 0.5.}
    \label{tab:cohen}
\end{table}

\section{Harmonic Mean Formulation}\label{app:harmonic_mean}
To evaluate the balance between recall and other important metrics (coverage and novelty), we use the following two harmonic means:
\begin{align}
h_{\text{RC}}@k &=\frac{2\times \text{Recall}@k \times \text{Coverage}@k}{\text{Recall}@k+\text{Coverage}@k},\\
h_{\text{RN}}@k &=\frac{2\times \text{Recall}@k \times \text{Novelty}@k}{\text{Recall}@k+\text{Novelty}@k}, 
\end{align}
where $h_{\text{RC}}@k$ represents the harmonic mean of Recall and Coverage at top-k recommendations, and $h_{\text{RN}}@k$ represents the harmonic mean of Recall and Novelty at top-k recommendations.
These harmonic means provide a balanced measure of the model's performance, considering both Recall and their diversity (i.e., Coverage and Novelty). By using these metrics, we can evaluate how well models balance these often competing objectives in recommendation systems.

\section{Full Results for Robustness to Noise Interactions}\label{app:noise}
We extend the results in the main content to evaluate the robustness of RDGCL and the baseline models to noise in user-item interactions and report on all metrics.

Tables~\ref{tab:noise-yelp} and~\ref{tab:noise-gowalla} show the performance of the models with respect to the noise ratio.
Our analysis reveals that RDGCL consistently outperforms all baseline models across different noise ratios (0.1\%, 0.3\%, 0.5\%) on both datasets, showing its robustness for noisy interactions.  As the noise ratio increases, we observe a general trend of performance degradation across all models, but RDGCL shows the least decline, indicating its superior noise resistance. Interestingly, on Yelp, the performance gap between RDGCL and SimGCL narrows slightly as noise increases, but RDGCL maintains its superiority.

On Gowalla, SGL appears most sensitive to noise, with significant performance drops as noise increases. SimGCL and LightGCL show intermediate robustness, but still underperform compared to our RDGCL.

\begin{table}[h]
    \small
    \setlength{\tabcolsep}{1.8pt}
    \centering
    \begin{tabular}{ll ccccc}\toprule
        Noise &  Metric     & LightGCN & SGL   & SimGCL  & LightGCL & RDGCL \\\midrule
        \multirow{4}{*}{\rotatebox[origin=c]{0}{0.1\%}}
                &  Recall@20 & 0.0762 & 0.0229 & 0.0964 & 0.0912 & 0.1075 \\
                &  NDCG@20   & 0.0647 & 0.0191 & 0.0815 & 0.0752 & 0.0917 \\
                &  Recall@40 & 0.1249 & 0.0550 & 0.1543 & 0.1467 & 0.1691 \\
                &  NDCG@40   & 0.0827 & 0.0313 & 0.1026 & 0.0957 & 0.1141 \\\midrule
        \multirow{4}{*}{\rotatebox[origin=c]{0}{0.3\%}}
                &  Recall@20 & 0.0749 & 0.0264 & 0.0955 & 0.0788 & 0.1023 \\
                &  NDCG@20   & 0.0633 & 0.0215 & 0.0808 & 0.0643 & 0.0866 \\
                &  Recall@40 & 0.1235 & 0.0417 & 0.1530 & 0.1310 & 0.1614 \\
                &  NDCG@40   & 0.0812 & 0.0270 & 0.1016 & 0.0837 & 0.1083 \\\midrule
        \multirow{4}{*}{\rotatebox[origin=c]{0}{0.5\%}}
                &  Recall@20 & 0.0741 & 0.0278  & 0.0953  & 0.0709 & 0.0987\\
                &  NDCG@20   & 0.0620 & 0.0218  & 0.0800  & 0.0580 & 0.0831\\
                &  Recall@40 & 0.1223 & 0.0461  & 0.1532  & 0.1187 & 0.1573\\
                &  NDCG@40   & 0.0797 & 0.0286  & 0.1011  & 0.0757 & 0.1045\\\bottomrule
    \end{tabular}
    \caption{The results w.r.t. noise ratio on Yelp}
    \label{tab:noise-yelp}
\end{table}

\begin{table}[h]
    \small
    \setlength{\tabcolsep}{1.8pt}
    \centering
    \begin{tabular}{ll ccccc}\toprule
        Noise &  Metric     & LightGCN & SGL   & SimGCL  & LightGCL & RDGCL \\\midrule
        \multirow{4}{*}{\rotatebox[origin=c]{0}{0.1\%}}
                &  Recall@20 & 0.1329 &	0.1389 & 0.2363 & 0.2314 & 0.2535 \\
                &  NDCG@20   & 0.0803 &	0.0568 & 0.1400 & 0.1360 & 0.1526 \\
                &  Recall@40 & 0.1921 &	0.2412 & 0.3258 & 0.3222 & 0.3432 \\
                &  NDCG@40   & 0.0957 &	0.0843 & 0.1635 & 0.1596 & 0.1761 \\\midrule
        \multirow{4}{*}{\rotatebox[origin=c]{0}{0.3\%}}
                &  Recall@20 & 0.1310 & 0.0719 & 0.2352 & 0.2243 & 0.2487 \\
                &  NDCG@20   & 0.0793 & 0.0317 & 0.1381 & 0.1314 & 0.1483 \\
                &  Recall@40 & 0.1903 & 0.1426 & 0.3253 & 0.3109 & 0.3380 \\
                &  NDCG@40   & 0.0948 & 0.0496 & 0.1617 & 0.1541 & 0.1719 \\\midrule
        \multirow{4}{*}{\rotatebox[origin=c]{0}{0.5\%}}
                &  Recall@20 & 0.1305 & 0.0581 & 0.2346 & 0.2144 & 0.2435 \\
                &  NDCG@20   & 0.0789 & 0.0270 & 0.1369 & 0.1256 & 0.1435 \\
                &  Recall@40 & 0.1878 & 0.1097 & 0.3225 & 0.3001 & 0.3326 \\
                &  NDCG@40   & 0.0939 & 0.0399 & 0.1601 & 0.1481 & 0.1671\\\bottomrule
    \end{tabular}
    \caption{The results w.r.t. noise ratio on Gowalla}
    \label{tab:noise-gowalla}
\end{table}

\section{Sensitivity Analyses}\label{app:sens}
\subsubsection{Sensitivity to $\tau$}
We test our model for various settings of $\tau$, and the results are shown in Fig.~\ref{fig:sens_temp}. In Gowalla, performance improves as the value of $\tau$ increases until reaching an optimal point around 0.4. As $\tau$ becomes too large (e.g., $\tau \geq 0.5$), the performance decreases drastically in both datasets.

\subsubsection{Sensitivity to $\lambda_1$}
\begin{figure}[h]
    \centering
    \subfigure[Yelp]{\includegraphics[width=0.45\columnwidth]{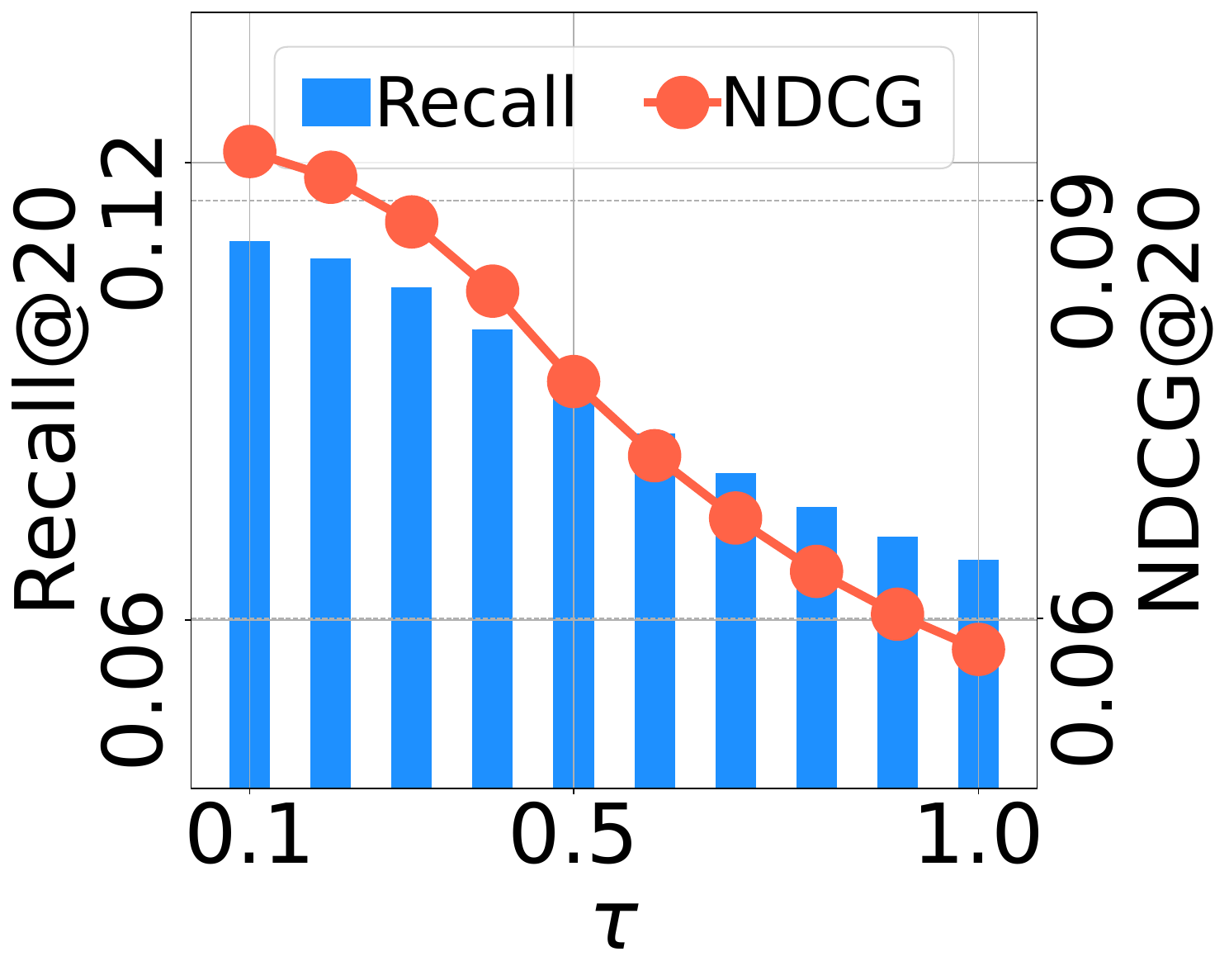}}
    \subfigure[Gowalla]{\includegraphics[width=0.45\columnwidth]{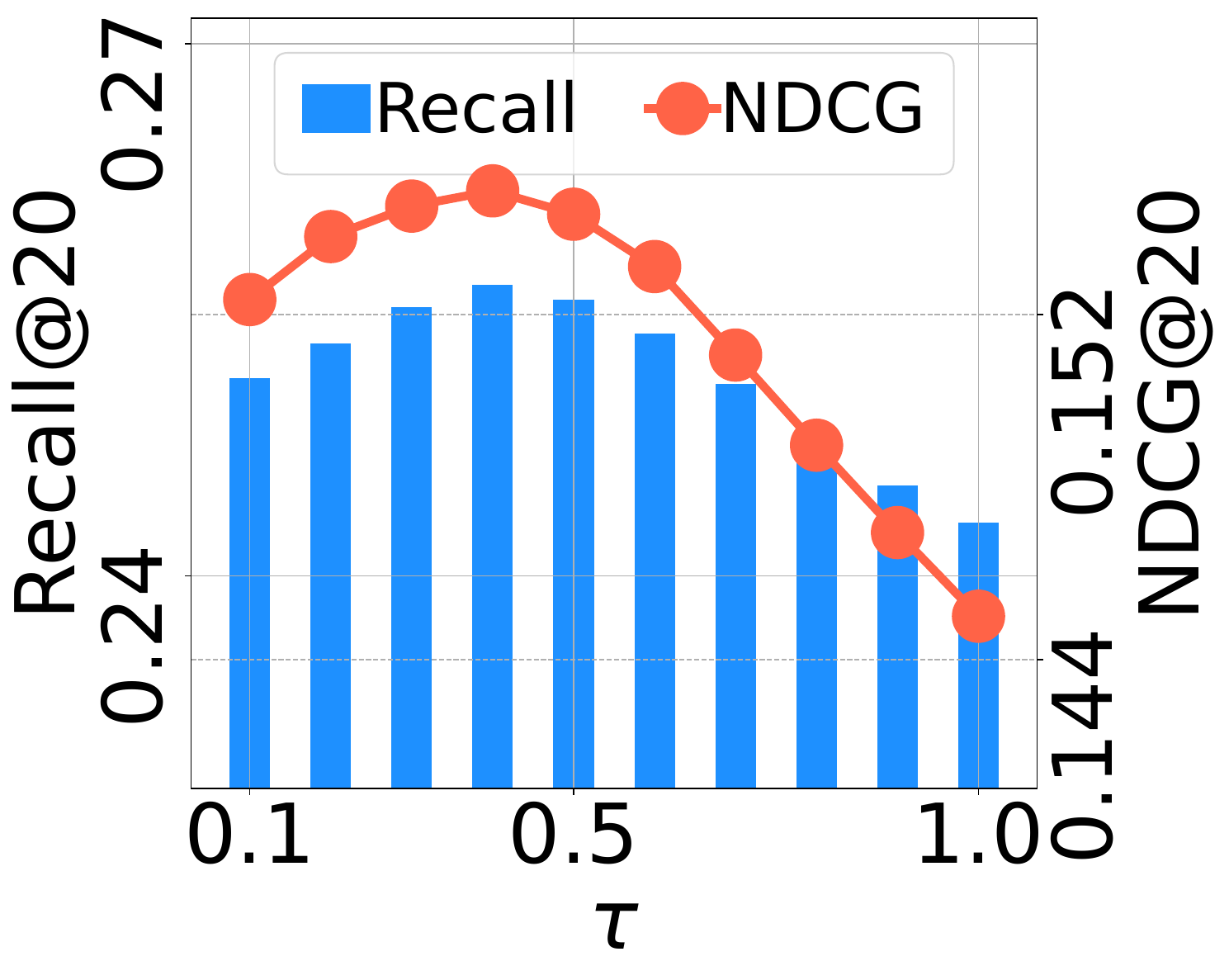}}
    \caption{Sensitivity on $\tau$}
    \label{fig:sens_temp}
\end{figure}

\begin{figure}[h]
    \centering
    \subfigure[Yelp]{\includegraphics[width=0.45\columnwidth]{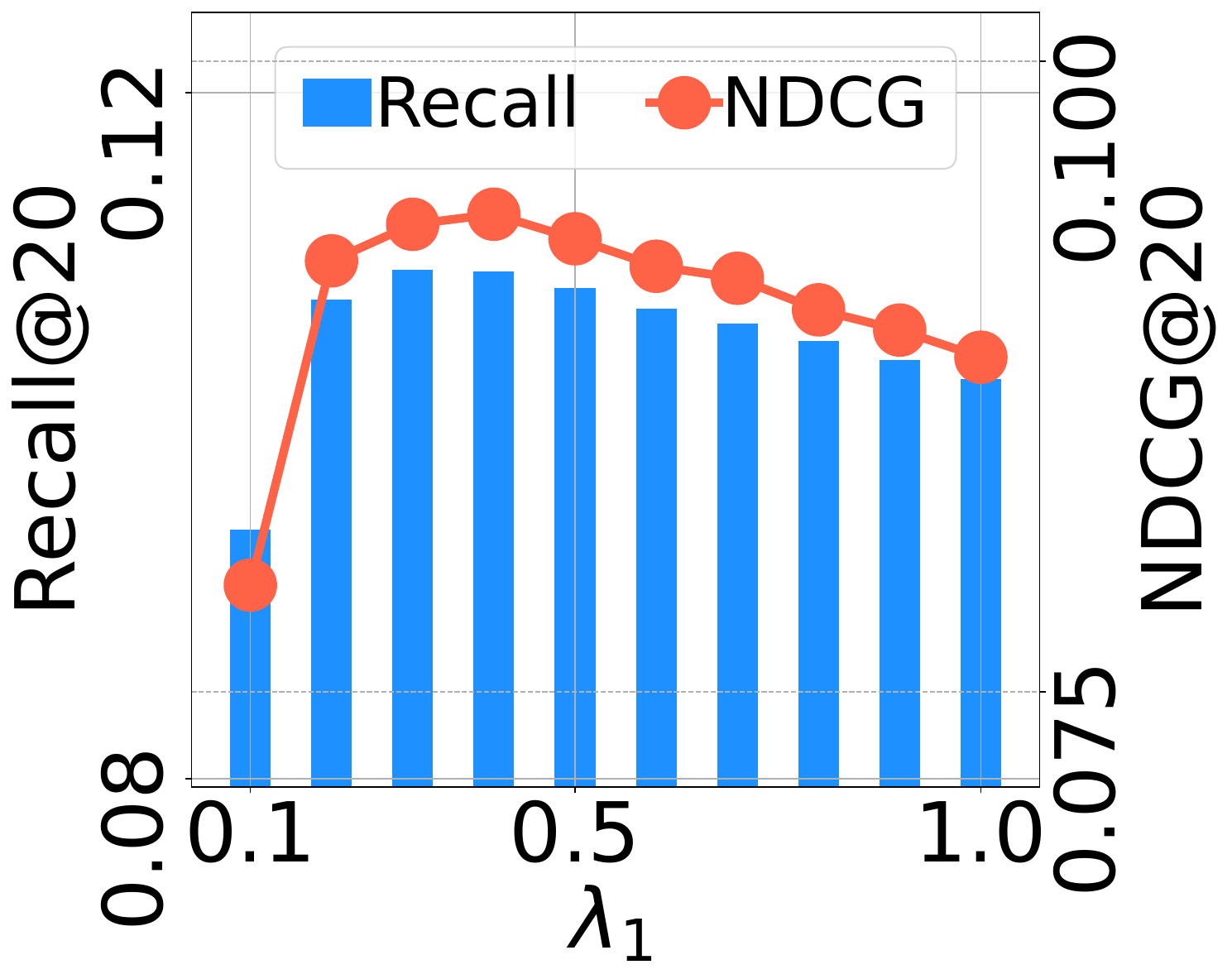}}
    \subfigure[Gowalla]{\includegraphics[width=0.45\columnwidth]{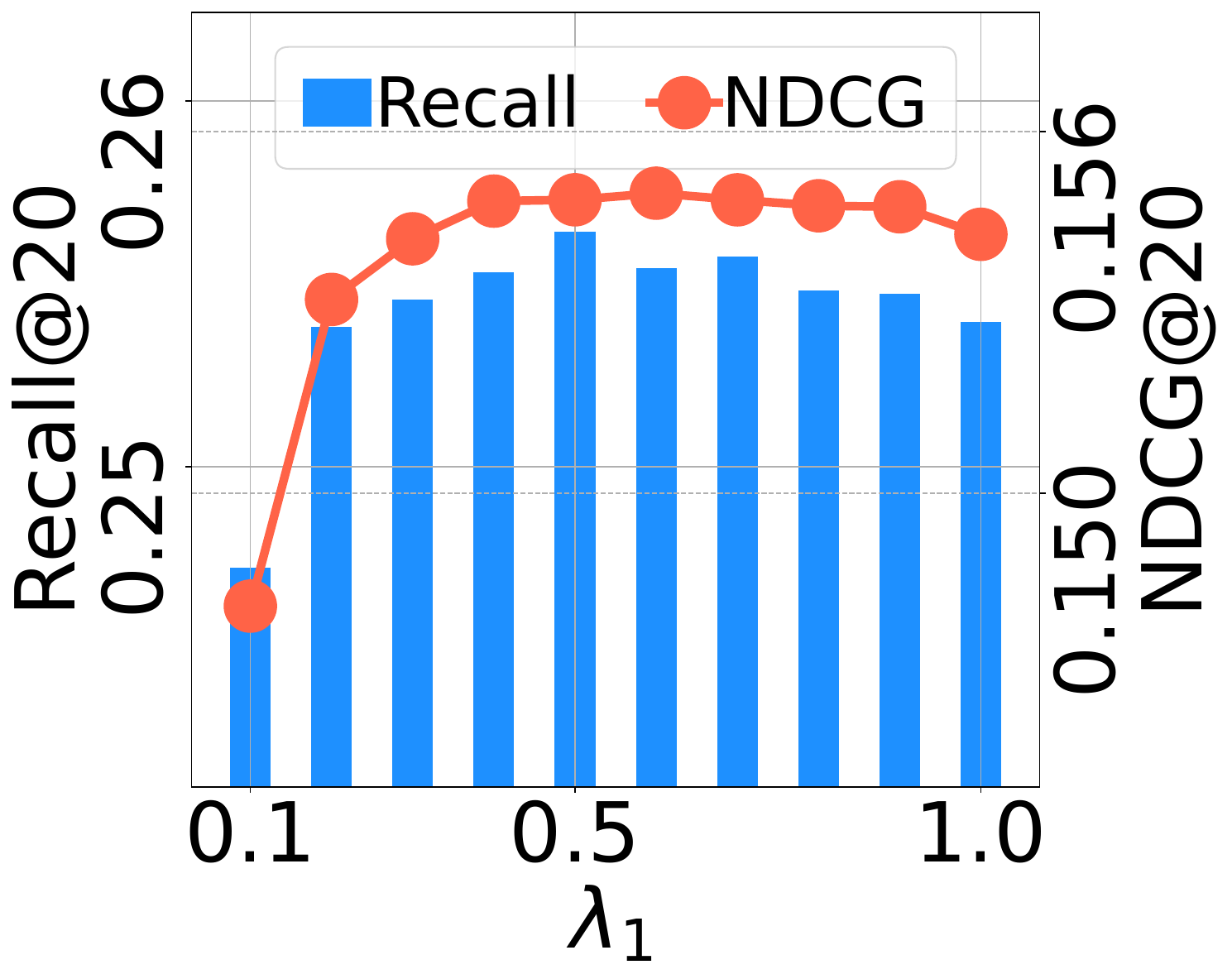}}
    \caption{Sensitivity on $\lambda_1$}
    \label{fig:sens_lambda1}
\end{figure}

We vary the regularization weight for the InfoNCE loss, denoted $\lambda_1$. The results are shown in Fig.~\ref{fig:sens_lambda1}. For both Yelp and Gowalla, the performance of our RDGCL increases rapidly as $\lambda_1$ increases. After that, it shows slight decreases. However, the performance does not decrease as much as that of Fig.~\ref{fig:sens_temp}.

\end{document}